%% file: Tensor_GSP.tex
\documentclass{IEEEtran}
\usepackage{algorithm,algorithmic,amsmath,amsfonts,amssymb,amsthm,cite,graphicx,epstopdf,threeparttable,multirow,pifont,stmaryrd,hyperref,dsfont,balance}
\usepackage[tight,footnotesize]{subfigure}
\usepackage[usenames]{color}
\usepackage[normalem]{ulem}
\IEEEoverridecommandlockouts

\input{_defs}

\theoremstyle{plain}

\newtheorem{theorem}{Theorem}

\theoremstyle{definition}

\theoremstyle{remark}
\newtheorem{remark}{Remark}

\graphicspath{{./}}
\begin{document}

\title{Learning Product Graphs Underlying\\ Smooth Graph Signals}

% authors
\author{Muhammad Asad Lodhi and Waheed U. Bajwa% <-this % stops a space
% thanks
\thanks{MAL and WUB are with the Department of Electrical and Computer Engineering, Rutgers, The State University of New Jersey, Piscataway, NJ 08854 (Emails: {\tt \{masad.lodhi,~waheed.bajwa\}@rutgers.edu}).}\thanks{This research is supported in part by the National Science Foundation under grants CCF-1453073 and CCF-1910110, and by the Army Research Office under grant W911NF-17-1-0546.}}

% running headings
%\markboth{}{}

\maketitle

%***************
\begin{abstract}
Real-world data is often times associated with irregular structures that can analytically be represented as graphs. Having access to this graph, which is sometimes trivially evident from domain knowledge, provides a better representation of the data and facilitates various information processing tasks. However, in cases where the underlying graph is unavailable, it needs to be learned from the data itself for data representation, data processing and inference purposes. Existing literature on learning graphs from data has mostly considered arbitrary graphs, whereas the graphs generating real-world data tend to have additional structure that can be incorporated in the graph learning procedure. Structure-aware graph learning methods require learning fewer parameters and have the potential to reduce computational, memory and sample complexities.
In light of this, the focus of this paper is to devise a method to learn structured graphs from data that are given in the form of product graphs. Product graphs arise naturally in many real-world datasets and provide an efficient and compact representation of large-scale graphs through several smaller factor graphs. To this end, first the graph learning problem is posed as a linear program, which (on average) outperforms the state-of-the-art graph learning algorithms. This formulation is of independent interest itself as it shows that graph learning is possible through a simple linear program. Afterwards, an alternating minimization-based algorithm aimed at learning various types of product graphs is proposed, and local convergence guarantees to the true solution are established for this algorithm. Finally the performance gains, reduced sample complexity, and inference capabilities of the proposed algorithm over existing methods are also validated through numerical simulations on synthetic and real datasets.
\end{abstract}

%***************

%***************
\input{intro}
\input{prob_problem_form}
\input{learning_algo}
%
\input{numerical_experiments}
\input{conclusion}
\input{appendix}
%\begin{definition}[Super Cool Definition]
%A definition could not be cooler!
%\end{definition}

%***************
%\newpage
\balance

\end{document}

%% file: _defs.tex
\newcommand{\bA}        {\mathbf{A}}

\newcommand{\bb}        {\mathbf{b}}
\newcommand{\bB}        {\mathbf{B}}

\newcommand{\wtbb}      {\widetilde{\bb}}

\newcommand{\bc}        {\mathbf{c}}

\newcommand{\bd}        {\mathbf{d}}
\newcommand{\bD}        {\mathbf{D}}

\newcommand{\whbD}      {\widehat{\bD}}

\newcommand{\E}         {\mathbb{E}}

\newcommand{\be}        {\mathbf{e}}

\newcommand{\bI}        {\mathbf{I}}

\newcommand{\bL}        {\mathbf{L}}
\newcommand{\cL}        {\mathcal{L}}

\newcommand{\whbL}      {\widehat{\bL}}

\newcommand{\bM}        {\mathbf{M}}

\newcommand{\cN}        {\mathcal{N}}

\newcommand{\rmO}       {\mathrm{O}}

\newcommand{\cO}        {\mathcal{O}}

\newcommand{\bp}        {\mathbf{p}}
\newcommand{\bP}        {\mathbf{P}}

\newcommand{\bQ}        {\mathbf{Q}}

\newcommand{\R}         {\mathbb{R}}

\newcommand{\bs}        {\mathbf{s}}
\newcommand{\bS}        {\mathbf{S}}

\newcommand{\wtbs}      {\widetilde{\bs}}
\newcommand{\wtbS}      {\widetilde{\bS}}

\newcommand{\bT}        {\mathbf{T}}
\newcommand{\cT}        {\mathcal{T}}

\newcommand{\bU}        {\mathbf{U}}

\newcommand{\bw}        {\mathbf{w}}
\newcommand{\bW}        {\mathbf{W}}
\newcommand{\cW}        {\mathcal{W}}

\newcommand{\whbw}      {\widehat{\bw}}
\newcommand{\whbW}      {\widehat{\bW}}

\newcommand{\bx}        {\mathbf{x}}
\newcommand{\bX}        {\mathbf{X}}
\newcommand{\cX}        {\mathcal{X}}

\newcommand{\by}        {\mathbf{y}}

\newcommand{\cY}        {\mathcal{Y}}

\newcommand{\bz}        {\mathbf{z}}
\newcommand{\bZ}        {\mathbf{Z}}

\newcommand{\bDelta}    {\boldsymbol{\Delta}}

\newcommand{\whbDelta}  {\widehat{\bDelta}}
\newcommand{\bLambda}   {\boldsymbol{\Lambda}}

\newcommand{\bzero}     {\mathbf{0}}
\newcommand{\bone}      {\mathbf{1}}
\newcommand{\one}      {\mathds{1}}
\DeclareMathOperator*{\argmin}  {arg\,min}
\DeclareMathOperator*{\argmax}  {arg\,max}
\newcommand{\diag}      {\mathrm{diag}}
\newcommand{\tr}        {\mathrm{trace}}

\newcommand{\tvec}      {\mathrm{vec}}

%% file: intro.tex
\section{Introduction} \label{sec:intro}
Graph signal processing (GSP) is an emerging field in data science and machine learning that aims to generalize existing information processing methods to data that live on an irregular domain. This underlying irregular domain can be represented as a graph and analysis of signals on the vertices of this graph, aptly named graph signals, is enabled by the graph shift operator (GSO). Recent developments in GSP have already established that GSO-based data processing outperforms classical signal processing for  several common tasks such as noise removal, signal filtering, wavelet representations, etc. \cite{bronstein2017geometric,sandryhaila2013discrete,sandryhaila2014big,shuman2013emerging,ortega2018graph}. The GSO is at the core of graph signal processing and could refer to either the adjacenecy matrix or one of the many types of Laplacian matrices associated with a graph. The exact choice of the GSO depends on the signal domain and the application of interest.
The eigenvectors of GSO provide bases for the spectral analysis of graph signals and generalize the concepts of bandlimited signals to the graph domain \cite{bronstein2017geometric,sandryhaila2013discrete,sandryhaila2014big,shuman2013emerging,ortega2018graph}. GSO also facilitates the synthesis of graph-based filters \cite{pasdeloup2017characterization,egilmez2018graph} %and is a key component in the description of rumor propagation over networks \cite{pasdeloup2017characterization,egilmez2018graph}. Moreover, GSO
and plays a pivotal role in the description of the notion of \emph{smoothness} for graph signals \cite{bronstein2017geometric,sandryhaila2013discrete,sandryhaila2014big,shuman2013emerging,ortega2018graph}.% and thus enables graph-based denoising strategies.

The underlying graph (and hence the GSO) for some real-world datasets is either known apriori, or can trivially be constructed through domain knowledge. As an example, consider weather data collected over a region. In this example, different weather stations would act as nodes, their observations as graph signals, and one (possible) way to construct the graph would be to connect physically adjacent nodes.
%This data consists of a set of weather stations distributed geographically over the region, where each weather station provides measurements of some atmospheric quantities (e.g., temperature, humidity, precipitation, etc.) over a period of time. One (trivial) way to construct the graph underlying this data would be to consider each weather station in this example as a vertex of the graph, with the length of an edge between any two vertices represented by the distance between them.
For most real-world data, however, such a trivially constructed graph is either non-optimal or it cannot be constructed in the first place due to lack of precise knowledge about the data generation process. This presents the need to learn the true underlying graph from the data itself. In this regard, the problem of graph learning from the observed data (i.e., graph signals) has gained a lot of attention in the recent years \cite{pasdeloup2017characterization,segarra2017network,egilmez2018graph,chepuri2017learning,kalofolias2017large,egilmez2017graph,dong2016learning,bronstein2017geometric}.

Graph learning refers to the problem of learning an unknown underlying graph from observed graph signals by exploiting some property of the graph signals. Traditional approaches for graph learning have proposed algorithms whose best-case complexity scales quadratically with the number of nodes in the graph \cite{dong2016learning,kalofolias2017large,kalofolias2016learn,chepuri2017learning,egilmez2017graph}. These approaches might be suitable for learning small graphs, but even for moderately sized graphs the learning cost would be prohibitive. Moreover, for learning an arbitrary graph (Laplacian), the number of parameters one needs to learn also scales quadratically with the number of nodes. Both of these problems hinder the amenability of traditional graph learning approaches to large-scale real-world datasets. Our work on graph learning, in contrast, hinges on the fact that real-world data is often generated over graphs that have an additional inherent structure. This inherent structure is dictated by either the way the data is acquired, by the arrangement of the sensors, or by the inherent relation of variables being observed \cite{sandryhaila2014big}. %(see Sec.~\ref{sec:product_graphs} for examples of various structures considered).
Moreover, this inherent structure of the graph being considered also presents itself in the associated GSO, which can incidentally be represented in terms of the product of several smaller \emph{factor} GSOs. In this paper, we will focus on three such structures that can be described in terms of three different products termed Kronecker, Cartesiasn and strong products.  Although aware of the presence of these product structures in real-world graphs (and the associated GSOs) \cite{sandryhaila2014big}, the research community has yet to propose algorithms that incorporate the graph product structure in the graph learning procedure.\footnote{During the course of revising this paper, we became aware of a recent work~\cite{kadambari2020learning} that aims to learn a Cartesian structured graph from data. This work targets a different objective function than ours, and (additionally) our method also addresses the problems of learning Kronecker and strong graphs, both of which fall outside the scope of the method proposed in \cite{kadambari2020learning}.} Additionally, as the number of free parameters scales quadratically with the number of nodes in the graph, and given the massive nature of the datasets available today, it has become imperative to devise methods that fully utilize the product structure of graphs to reduce the number of parameters to be learned. Moreover, posing the problems in terms of smaller factor graphs instead of the graph itself can enable efficient data representation \cite{sandryhaila2014big}, and can result in reduced sample complexity as one has to learn fewer parameters. %, and offer computational gains towards scalable processing and inference from huge datasets (see Sec.\ref{sec:numerical_experiments}).
To this end, the main objective of this work is to investigate the problem of learning product graphs from data in an efficient and scalable manner.

\subsection{Prior work} \label{sec:prior}
The existing works in graph signal processing can mainly be divided into four chronological categories.
The first set of works in GSP introduced the idea of information processing over graphs \cite{shuman2013emerging,sandryhaila2013discrete,ortega2018graph}. These works highlight the advantages and superior performance of graph-based signal processing approach (with known underlying graph) over classical signal processing.
The second wave of research in this area built upon the first one to exploit knowledge of the underlying graph for graph signal recovery from samples obtained over a subset of graph nodes or from noisy observations of all nodes \cite{chen2015signal,chen2016signal}. Through these works, the idea of bandlimitidness was extended to graph signals and the concept of smooth graph signals was introduced.
Since the underlying graph is not always available beforehand, the third wave of GSP analyzed the problem of recovering the underlying graph through observations over the graph \cite{pasdeloup2017characterization,segarra2017network,egilmez2018graph,chepuri2017learning,kalofolias2017large,egilmez2017graph,dong2016learning}. Finally, the fourth wave of research in GSP has focused on joint identification of the underlying graph and graph signals from samples/noisy observations using interrelated properties of graph signals and graphs \cite{ioannidis2018semi,ceci2018signal,sardellitti2019graph,dong2016learning}.

Within the third set of papers in GSP, our work falls in the category of combinatorial graph Laplacian estimation \cite{chepuri2017learning,kalofolias2017large,egilmez2017graph,dong2016learning} from graph signals. Combinatorial graph Laplacian refers to the unnormalized Laplacian of an unstructured graph with no self loops \cite{egilmez2017graph}. The earliest of works in this category \cite{dong2016learning} aims to jointly denoise noisy graph signals observed at the nodes of the graph and also learn the graph from these denoised graph signals. The authors pose this problem as a multiconvex problem in the graph Laplacian and the denoised graph signals, and then solve it via an alternating minimization approach that solves a convex problem in each unknown. The authors in \cite{segarra2017network} examine the problem of graph Laplacian learning when the eigenvectors of the graph Laplacian are known beforehand. They achieve this by formulating a convex program to learn a valid graph Laplacian (from a feasible set) that is diagonalized by the noiseless and noisy Laplacian eigenvectors.

The work in \cite{chepuri2017learning} takes a slightly different route and learns a sparse unweighted graph Laplacian matrix from noisy graph signals through an alternating minimization approach that restricts the number of edges in the graph. In contrast to earlier work, \cite{pasdeloup2017characterization} focuses on learning a graph diffusion process from observations of stationary signals on graphs through convex formulations.  In this regard, the authors also consider different criteria in addition to searching for a valid Laplacian and devise specialized algorithms to infer the diffusion processes under these criteria. In \cite{kalofolias2016learn,kalofolias2017large} the graph learning problem is adressed by posing it in terms of learning a sparse weighted adjacency matrix from the observed graph signals. Finally, the authors in \cite{egilmez2017graph} provide a comprehensive unifying framework for inferring several types of graph Laplacians from graph signals. They also make connection with the state-of-the-art and describe where the past works fit in light of their proposed framework.

It should be mentioned here that since Laplacian matrices (and thus adjacency matrices) are related to precision matrices, defined as the (pseudo-)inverses of covariance matrices, imposing a structure on the graph adjacency matrix amounts to imposing a structure on the covariance of data. Earlier works in the field have already made comparisons of Laplacian learning approaches with those for learning precision matrices from data \cite{dong2016learning,egilmez2017graph}, and established the superior graph recovery performance of Laplacian-based learning. Some recent works have also investigated learning structured covariance and precision matrices, and their usefulness in efficiently representing real-world datasets \cite{friedman2008sparse,greenewald2017tensor,tsiligkaridis2013covariance}. While these models work well in practice, we will demonstrate through our experiments that there are scenarios where graph-based learning outperforms structured covariance-based learning (see Sec.~\ref{sec:numerical_experiments} for details).

\subsection{Our contributions} \label{sec:our}
Our first contribution in this work is a novel formulation of the graph learning problem as a linear program. We show, both theoretically and empirically, that graph adjacency matrices can be learned through a simple and fast linear program. We then shift our attention towards learning structured graphs. Most of the prior works regarding graph learning have considered arbitrary graphs with either some connectivity constraints \cite{kalofolias2017large}, or no constraints at all \cite{egilmez2018graph,dong2016learning}. In all cases, the complexity of the graph learning procedure and the number of free parameters scale quadratically with the number of nodes in the graph, which can be prohibitively large in real-world scenarios. In contrast, our work focuses on inferring the underlying graph from graph signals in the context of structured graphs. Specifically, we investigate graphs that can be represented as Kronecker, Cartesian, and strong products of several smaller graphs. We first show how, for these product graphs, the graph adjacency matrix, the graph Laplacian, the graph Fourier transform, and the graph smoothness measure can be represented with far fewer parameters than required for arbitrary graphs. This reduction in number of parameters to be learned results in reduced sample complexity and helps avoid overfitting. Afterwards, we outline an algorithm to learn these product graphs from the data and provide convergence guarantees for the proposed algorithm in terms of the estimation error of factor graphs. We validate the performance of our algorithm with numerical experiments on both synthetic and real data.
%Finally, in contrast to existing efforts in graph learning, we provide an empirical procedure to select the best candidate graph from the set of candidate graphs learned over different values of our optimization parameters.

\subsection{Notation and organization}
The following notational convention is used throughout the rest of this paper. We use bold lower-case and bold-upper case letters to represent vectors and matrices, respectively. Calligraphic letters are used to represent tensors, which are arrays of three or more dimensions. For a tensor $\cT$, $\cT_{(i)}$ represents its matricization (flattening) in the $i$-th mode and $\tvec(\cT)$ represents its vectorization along the first mode \cite{kolda2009tensor}. Also, ``$:$" represents the scalar product or double dot product between two tensors, which results in a scalar \cite{kolda2009tensor}. The Hadamard product (elemntwise product) of two vectors or matrices is denoted by ``$\circ$". For a matrix $\bA$, $\|\bA\|_{F}$ represents its Frobenius norm, $\|\bA\|$ represents its spectral norm, and $\bA^\dagger$ represents its Moore-Penrose inverse. Moreover, $\|\bA\|_{1}$ represents the $\ell_{1}$-norm of the entries of $\bA$, while $\|\bA\|_{1,\text{off}}$ represents the $\ell_{1}$-norm of the off-diagonal entries of $\bA$. The Kronecker and Cartesian products of two matrices $\bA$ and $\bB$ are denoted by $\bA \otimes \bB$ and $\bA \oplus \bB$, respectively \cite{horn1991topics}. The strong product of two matrices is denoted by $ \bA \boxtimes \bB = \bA \otimes \bB ~+~ \bA \oplus \bB$, which is the sum of Cartesian and Kronecker products of the respective matrices. Furthermore, $\underset{\bs}{\otimes}$, $\underset{\bs}{\oplus}$, and $\underset{\bs}{\boxtimes}$, respectively denote the Kronecker, Cartesian, and strong products taken over the indices provided by the entries of the vector $\bs$. Finally, $\times_{i}$ represents matrix multiplication in the $i$-th mode of a tensor and $\underset{\bs}{\times}$ represents matrix multiplications in the modes of a tensor specified in the entries of the vector $\bs$.

We use $\diag(\bx)$ to denote a diagonal matrix with diagonal entries given by the entries of the vector $\bx$, $\bone$ to denote a vector of all ones with appropriate length, and $\mathds{1}$ to denote a tensor of all ones of appropriate size. We denote the set with elements $\{1,2,\dots,K\}$ as $[K]$, and $[K] \setminus k$ represents the same set without the element $k$. The sets of valid Laplacian and weighted adjacency matrices are represented by $\cL$ and $\cW$, respectively. The set of weighted adjacency matrices with any product structure is denoted by $\cW_{p}$.

{The rest of this paper is organized as follows. In Sec.~\ref{sec:prob_problem_form} we give a probabilistic formulation of the graph learning problem in line with existing literature. Then, we propose our novel formulation of the graph learning problem as a liner program in Sec.~\ref{sec:graph_learning_linear}. Sec.~\ref{sec:product_graphs} describes the motivation for product graphs and formulates the graph learning problem in the context of product graphs. In Sec.~\ref{sec:learning_algo} we propose an algorithm for learning product graphs from data and derive error bounds on estimated factor graphs. We present our numerical experiments with synthetic and real datasets in Sec.~\ref{sec:numerical_experiments}, and the paper is concluded in Sec.~\ref{sec:conc}.} 

%% file: prob_problem_form.tex
\section{Probabilistic Problem formulation} \label{sec:prob_problem_form}
In this section we formulate the arbitrary graph learning problem from a probabilistic standpoint. Let us assume access to $m = 1,\dots,M_{0}$ graph signals $\bx_{m} \in \R^n$ observed on $n$ nodes of an undirected graph $G = \{V, E\}$ without any self loops, where $V$ and $E$ represent the nodes and edges of the graph. Weighted edges of this graph can be represented as a weighted adjacency matrix $\bW \in \R^{n \times n}$, which has a zero diagonal owing to no self-loops in the graph. Based on the adjacency matrix $\bW$, one can define the degree matrix $\bD = \diag(W \bone)$, which is a diagonal matrix containing the weighted degree of each node at the respective diagonal entry. The associated unnormalized graph Laplacian can then be defined as $\bL = \bD - \bW$. The adjacency matrix $\bW$ of the graph can be decomposed as $\bU \bLambda \bU^{T}$ and its eigenvectors define the graph Fourier basis for the graph Fourier transform \cite{sandryhaila2014big}.

The signals observed on the nodes of a graph are assumed to have a joint distribution given by a multivariate normal distribution, i.e., $\bx_{m} \sim \cN(0, \bL^{\dagger})$, where $\bL^{\dagger}$ is the pseudoinverse of $\bL$ and $\bL$ represents the graph Laplacian. In words, signals generated over a graph can be seen as being generated over a Gaussian Markov Random Field (GMRF) whose precision matrix is the graph Laplacian \cite{dong2016learning}. Given independent observations $\{\bx_{m}\}$, the maximum likelihood estimate (MLE) of $\bL$ can be expressed as:
\begin{align} \label{learn_lap}
\widehat{\bL}_{\text{MLE}} &= \underset{\bL \in \cL}{\argmax}  |\bL|^{\frac{M_{0}}{2}} \exp(-\frac{1}{2} \underset{m = 1}{\overset{M_{0}}{\sum}} \bx_{m}^{T} \bL \bx_{m}) \nonumber\\
&= \underset{\bL \in \cL}{\argmin}  -\log|\bL| + \frac{1}{M_{0}}~\underset{m = 1}{\overset{M_{0}}{\sum}} \bx_{m}^{T} \bL \bx_{m},
\end{align}
where $\cL$ represents the class of valid Laplacians, i.e., a symmetric positive semi-definite matrix with rows that sum to zero and nonpositive off-diagonal entries. With the Laplacian constraints, the problem in \eqref{learn_lap} can be further expressed as:
\begin{align} \label{learn_lap_2}
&\widehat{\bL}_{\text{MLE}} =~\underset{\bL}{\argmin} ~ -\log|\bL| + \frac{1}{M_{0}}~\underset{m = 1}{\overset{M_{0}}{\sum}} \bx_{m}^{T} \bL \bx_{m} \nonumber\\
&\quad\quad\quad~\text{s.t.  } 	\bL \bone = \bzero,~ \tr (\bL) = n,~(\bL)_{ij} = (\bL)_{ji} \leq 0.
\end{align}

Interactions in the real world tend to be mostly local, and thus not all nodes in a graph are connected to each other in real-world datasets. To impose only local interactions, usually a sparsity term regularizing the off-diagonal entries of the Laplacian matrix is added to the graph learning objective to learn \textit{sparse} graphs. Therefore, traditional graph learning approaches \cite{chepuri2017learning,kalofolias2017large,egilmez2017graph,dong2016learning,kalofolias2016learn} take a form similar to the following:
\begin{align} \label{learn_lap_sparse}
&\widehat{\bL}_{\text{REG}} =~\underset{\bL}{\argmin} ~ -\log|\bL| + \frac{\alpha}{M_{0}}~\underset{m = 1}{\overset{M_{0}}{\sum}} \bx_{m}^{T} \bL \bx_{m} + \beta \|\bL\|_{1,\text{off}} \nonumber\\
&\quad\quad\quad~\text{s.t.  } 	\bL \bone = \bzero,~ \tr (\bL) = n,~(\bL)_{ij} = (\bL)_{ji} \leq 0,
\end{align}
where $\|\bL\|_{1,\text{off}}$ represents a sparsity penalty on the off-diagonal entries of $\bL$, and parameters $\alpha$ and $\beta$ control the penalty on the quadratic term and the density of the graph, respectively.
In the following section, we show that the traditional graph learning problem can be significantly simplified and that %arbitrary 
graphs can actually be learned through a simple linear program.

%\section{Graph learning as a linear program} \label{sec:graph_learning_linear}
%In this section we propose linear formulations of the graph learning problems for arbitrary and structured graphs. We first formulate the linear problem for arbitrary graphs and then specialize this problem for learning Kronecker, Cartesian and strong product graphs from data.
%
%\subsection{Linear program for arbitrary graphs} \label{sec:graph_learning_linear:arbitrary}
\section{Graph learning as a linear program} \label{sec:graph_learning_linear}
Let us start by inspecting the traditional graph learning problem in \eqref{learn_lap_sparse}.  In particular, let us first focus on the term $\log|\bL|$ in the objective function and the constraint $\tr (\bL) = n$. We can express this log-determinant term in the objective as $\log|\bL| = \sum_{i=1}^{n} \log \lambda_{i}$, where $\lambda_{i}$ is the $i$-th largest eigenvalue of $\bL$. Thus, through this $\log|\bL|$ term, \eqref{learn_lap_sparse} constrains the spectrum of the Laplacian matrix to be estimated. %This regularization is necessary for precision matrix estimation in GMRFs to avoid the trivial solution of an all zeros matrix. 
However, for our problem of estimating the Laplacian matrix, the constraint $\tr (\bL) = \sum_{i=1}^{n} \lambda_{i} = n$, is already putting a hard constraint on the sum of eigenvalues of the Laplacian matrix. Moreover, the constraint $\bL \bone = \bzero$ is forcing the smallest eigenvalue of the Laplacian to be zero. In the presence of these constraints, the log-determinant regularization in the objective function is no longer necessary to arrive at a valid Laplacian matrix or to avoid trivial solutions. Another advantage of removing the log-determinant term is the massive savings in computational complexity as this term forces one to employ singular value decomposition at each step of the learning algorithm \cite{greenewald2017tensor,tsiligkaridis2013covariance}. %In contrast, the relevant constraints in the problem are convex (linear) and admit fast and efficient solutions.

Let us also examine the term $\sum_{m=1}^{M_{0}} \bx_{m}^{T} \bL \bx_{m}$ in the objective in \eqref{learn_lap}. This term comes from the likelihood of the observed signals with the Laplacian as the precision matrix, and also represents the sum of Dirichlet energy or ``smoothness" of the observed graph signals \cite{kalofolias2016learn,dong2016learning,egilmez2017graph}. It has been shown in the existing literature \cite{kalofolias2016learn} that this term can be expressed as a weighted sparsity regularization on the graph adjacency matrix as $\sum_{m=1}^{M_{0}} \bx_{m}^{T} \bL \bx_{m} = \tr(\bX^{T} \bL \bX) = \| \bW \circ \bZ\|_{1}$. Here $\bX$ is the data matrix with $\bx_{m}$ as the $m$-th column, and $\bZ$ is the matrix of pairwise distances between rows of $\bX$ such that $(i,j)$-th entry in $\bZ$ is the Euclidean distance between the $i$-th and $j$-th row of $\bX$. This implies that the sum of Dirichlet energy in the objective implicitly regularizes the sparsity of $\bW$ and thus controls the density of edges in the graph. Therefore, presence of this term in the objective eliminates the need to explicitly regularize the sparsity of the graph to be learned.

In light of the preceding discussion, we propose to solve the following linear program \cite{boyd2004convex} for learning graphs:
\begin{align} \label{learn_lap_3}
\widehat{\bL} =~&\underset{\bL}{\min} ~~~ \frac{\alpha}{M_{0}}~\underset{m = 1}{\overset{M_{0}}{\sum}} \bx_{m}^{T} \bL \bx_{m} \nonumber\\
&\text{s.t.  } 	\bL \bone = \bzero,~ \tr (\bL) = n,~(\bL)_{ij} = (\bL)_{ji} \leq 0.
\end{align}
where $\alpha$ is a regularization parameter that controls the smoothness of the graph signals and thus the sparsity of edges in the graph.
%The product structure of the product graphs cannot be directly exploited through the graph Laplacian. Instead, a more favorable formulation would be to learn the graph adjacency matrix, which can be explicitly represented in terms of the product structure. More specifically, $\bx_{m}^{T} \bL \bx_{m} = \bx_{m}^{T} (\bD - \bW) \bx_{m} = \bx_{m}^{T} \bD \bx_{m} - \bx_{m}^{T} \bW \bx_{m} = \bar{\bx}_{m}^{T} \bW \bone - \bx_{m}^{T} \bW \bx_{m}$. With this, the final form of our graph learning problem becomes:
%\begin{align} \label{learn_adjacency}
%&\underset{\bW}{\argmin} &&\frac{\alpha}{M_{0}}~\underset{m = 1}{\overset{M_{0}}{\sum}} (\bar{\bx}_{m}^{T} \bW \bone - \bx_{m}^{T} \bW \bx_{m}) \nonumber\\
%&\quad\text{s.t. } 	&&\diag(\bW) = \bzero, \bone^{T} \bW \bone = n, (\bW)_{ij} = (\bW)_{ji} \geq 0.
%\end{align}
%where $\bar{\bx}_{m} = \bx_{m} \odot \bx_{m}$, and $\odot$ represents the entry-wise product of two vectors. This proposed formulation for learning graphs can be further specialized to the case of learning product graphs. We will discuss several cases of these product graphs in the following sections. 

\subsection{Fast solver for the graph learning linear program} \label{sec:learning_algo:linear_problem}
We now present an algorithm, named \textbf{G}raph learning with \textbf{L}inear \textbf{P}rogramming (GLP), for solving the linear graph learning problem \eqref{learn_lap_3}. %Our formulation of the graph learning problem proposed in the preceding sections is of independent interest as it can be used to learn general unstructured graphs (as pointed out in Sec.~\ref{sec:product_graph_prob:arbitrary}).
To proceed, note that the objective term in the graph learning problem can be reformulated as:
\begin{align} \label{smoothness_adjacency}
\frac{1}{M_{0}}\underset{m = 1}{\overset{M_{0}}{\sum}} &\bx_{m}^{T} \bL \bx_{m} = \frac{1}{M_{0}}\underset{m = 1}{\overset{M_{0}}{\sum}} (\bx_{m}^{T} \bD \bx_{m} - \bx_{m}^{T} \bW \bx_{m}) \nonumber\\
&= \frac{1}{M_{0}}\underset{m = 1}{\overset{M_{0}}{\sum}} (\bx_{m}^{T} \diag(\bW \bone) \bx_{m} - \bx_{m}^{T} \bW \bx_{m}) \nonumber\\
&= \frac{1}{M_{0}}\underset{m = 1}{\overset{M_{0}}{\sum}} (\bx_{m}^{T} \diag(\bx_{m} ) \bW \bone - \bx_{m}^{T} \bW \bx_{m}) \nonumber\\
&= \frac{1}{M_{0}}\underset{m = 1}{\overset{M_{0}}{\sum}} (\bar{\bx}_{m}^{T} \bW \bone - \bx_{m}^{T} \bW \bx_{m}) \nonumber\\
&= \frac{1}{M_{0}}\underset{m = 1}{\overset{M_{0}}{\sum}} \tr(\bar{\bx}_{m}^{T} \bW \bone - \bx_{m}^{T} \bW \bx_{m}) \nonumber\\
&= \frac{1}{M_{0}}\tr \Big[ \bW \underset{m = 1}{\overset{M_{0}}{\sum}} \bone \bar{\bx}_{m}^{T} - \bW \underset{m = 1}{\overset{M_{0}}{\sum}} \bx_{m} \bx_{m}^{T} \Big] \nonumber\\
&= \tr \Big[ \bW \Big(\underset{m = 1}{\overset{M_{0}}{\sum}} \bone \bar{\bx}_{m}^{T} - \underset{m = 1}{\overset{M_{0}}{\sum}} \bx_{m} \bx_{m}^{T} \Big)/M_{0} \Big] \nonumber\\
&= \tr (\bW \wtbS) = \tvec(\wtbS)^{T}\tvec(\bW) = \wtbs^{T} \bM \bw,
\end{align}
%\begin{align}
%&\alpha~\tr( \bD \bar{\bS}) - \alpha \tr( \bW \bS) \nonumber\\
%&= \alpha~\tr( \bW \wtbS) = \alpha~\tvec(\wtbS)^{T}\tvec(\bW) = \alpha~\wtbs^{T} \bD \bw,
%\end{align}
where the matrix $\wtbS = (\underset{m = 1}{\overset{M_{0}}{\sum}} \bone \bar{\bx}_{m}^{T} - \underset{m = 1}{\overset{M_{0}}{\sum}} \bx_{m} \bx_{m}^{T} )/M_{0}$, the vector $\bar{\bx}_{m}^{T} = \bx_{m}^{T} \diag(\bx_{m} ) = \bx_{m}^{T} \circ \bx_{m}^{T}$, $\tvec(\bW) = \bM \bw$, $\bw$ is a vector of distinct elements of upper triangular part of the symmetric matrix $\bW$, and $\bM$ is the duplication matrix that duplicates the elements from $\bw$ to generate a vectorized version of $\bW$. With this rearrangement of the objective and $\bW$, our graph learning problem can be posed as:
\begin{align}
\widehat{\bw} =~&\underset{\bw}{\min}~\alpha~\wtbs^{T} \bM \bw \quad \text{s.t. } 	\bA \bw = \bb,~ \bw_{i} \geq 0, i \in F,
\end{align}
where $\bA$ is a matrix that represents the equality constraints from \eqref{learn_lap_3} in terms of equality constraints on $\bw$, $\bb = [\bzero^{T}, n]^{T}$, and $F$ is the set containing the indices of the off-diagonal elements in $\bw$. Once the solution $\whbw$ is obtained, it can be converted to the symmetric adjacency matrix $\whbW$, which can then be used to get $\whbL$.
%This optimization problem is a linear program as it has a linear objective function and linear constraints.

The standard way of solving a linear program with mixed (equality and inequality) constraints is through interior point methods whose complexity scales quadratically with the problem dimension \cite{boyd2004convex}. A better alternative is to deploy a first-order method whose per-iteration complexity is linear in the number of nonzero entries of $\bA$. However, a first-order method would exhibit slow convergence for a linear program because of the lack of smoothness and strong convexity in linear programs \cite{wang2017new}.
%due to the lack of properties like smoothness and strong convexity in a linear program, a slow convergence rate of $\cO(1/\epsilon^{2})$ is achieved. 
To overcome these issues, linear programs have been solved through the Alternating Direction Method of Multiplier (ADMM) \cite{eckstein1990alternating,wang2017new}. To solve our proposed linear formulation of graph learning, we follow a recent algorithm proposed in \cite{wang2017new}. This ADMM-based algorithm for linear programs proposed a new variable splitting scheme that achieves a convergence rate of $\cO(\|\bA\|^2\log(1/\epsilon))$, where $\epsilon$ is the desired accuracy of the solution. To this end, we start by modifying the original graph learning problem with the introduction of an additional variable $\by$ as follows:
\begin{align}
\widehat{\bw} =~&\underset{\bw}{\min}~\bc^{T} \bw \quad \text{s.t. } 	\bA \bw = \bb,~ \bw = \by,~ \by_{i} \geq 0, i \in F,
\end{align}
where $\bc = \alpha \bM^{T} \wtbs$. The corresponding augmented Lagrangian can then be expressed as follows:
\begin{align} \label{eq:lagrangian}
L(\bw,\by,\bz) = \bc^{T} \bw + h(\by) + \bz^{T} (\bA_{\bw} \bw + \bA_{\by} \by - \wtbb) \nonumber\\
+ \rho/2 \| \bA_{\bw} \bw + \bA_{\by} \by - \wtbb \|_{2}^{2},
\end{align}
where $h(\by)$ denotes the non-negativity constraint on the entries of $\by$ indexed by $F$, i.e., $\forall i \in F$, $h(\by) = 0$ when $y_{i} \geq 0$, and $h(\by) = \infty$ when $y_{i} < 0$.
Moreover, $\bz = [\bz_{\bw}^{T}, \bz_{\by}^{T}]^{T}$, $\bz_{\bw}$ and $\bz_{\by}$ are the Lagrange multipliers, $\bA_{\bw} = [\bA^{T}, \bI]^{T}$, $\bA_{\by} = [\bzero, -\bI]^{T}$, and finally $\wtbb = [\bb^{T}, \bzero^{T}]^{T}$. One can then use ADMM to go through the %following steps until convergence:
steps outlined in Algorithm~\ref{alg_learn_graph} until convergence to obtain $\whbw$.
%\begin{algorithm}[t]
%	\caption{\textbf{: GLP---ADMM for graph learning with linear programming}}
%	\label{alg_learn_graph}
%	{\textbf{Input:}  Observations $\{\bx_{m}\}_{m = 1}^{M_{0}} $, maximum iterations $T_{0}$, and parameter $\alpha,\rho$}\\
%	\textbf{Initialize:} $\bw^{(1)} \leftarrow \bzero$,
%	\begin{algorithmic}
%		\STATE \textbf{for} $t=1$ to $T_{0}$
%		\STATE \hspace{\algorithmicindent} $\bw^{(t+1)} \leftarrow \rho^{-1} (\bI + \bA^{T}\bA)^{-1} \be^{(t)}$
%		\STATE \hspace{\algorithmicindent} $\by^{(t+1)} \leftarrow [\bw^{(t+1)} + \bz_{\by}^{(t)}/\rho]_{\underset{F}{\geqslant} \bzero}$
%		\STATE \hspace{\algorithmicindent} $\bz^{(t+1)} \leftarrow \bz^{(t)} + \rho(\bA_{\bw} \bw^{(t+1)} + \bA_{\by} \by^{(t+1)} - \wtbb)$
%		\STATE \hspace{\algorithmicindent} $\be^{(t+1)} \leftarrow - \bA_{\bw}^{T} [\bz^{(t+1)} + \rho(\bA_{\by} \by^{(t+1)} - \wtbb)] - \bc$
%		\STATE \textbf{end}
%	\end{algorithmic}
%	\textbf{Output:} Final adjacency estimate $\whbw \leftarrow \bw^{(t+1)}$.
%\end{algorithm}
\begin{algorithm}[t]
	\caption{\textbf{: GLP---ADMM for graph learning with linear programming}}
	\label{alg_learn_graph}
	{\textbf{Input:}  Observations $\{\bx_{m}\}_{m = 1}^{M_{0}} $, maximum iterations $T_{0}$, and parameters $\alpha,\rho > 0$}\\
	\textbf{Initialize:} $\by^{(1)} \leftarrow \bzero$~,~$\bz^{(1)} %= \begin{bmatrix} \bz_{\bw} \\ \bz_{\by} \end{bmatrix} 
																\leftarrow \bone$
	\begin{algorithmic}
		\STATE \textbf{for} $t=1$ to $T_{0}$
		\STATE \hspace{\algorithmicindent} $\be^{(t+1)} \leftarrow - \bA_{\bw}^{T} [\bz^{(t)} + \rho(\bA_{\by} \by^{(t)} - \wtbb)] - \bc$
		\STATE \hspace{\algorithmicindent} $\bw^{(t+1)} \leftarrow \rho^{-1} (\bI + \bA^{T}\bA)^{-1} \be^{(t+1)}$
		\STATE \hspace{\algorithmicindent} $\by^{(t+1)} \leftarrow [\bw^{(t+1)} + \bz_{\by}^{(t)}/\rho]_{\underset{F}{\geqslant} \bzero}$
		\STATE \hspace{\algorithmicindent} $\bz^{(t+1)} \leftarrow \bz^{(t)} + \rho(\bA_{\bw} \bw^{(t+1)} + \bA_{\by} \by^{(t+1)} - \wtbb)$
		\STATE \textbf{end}
	\end{algorithmic}
	\textbf{Output:} Final adjacency estimate $\whbw \leftarrow \bw^{(t+1)}$
\end{algorithm}

In Algorithm~\ref{alg_learn_graph}, $[\cdot]_{\underset{F}{\geqslant} \bzero}$ is entrywise thresholding that projects the entries with indices in $F$ to the nonnegative orthant. As we can see from the algorithm, all updates have closed-form solutions and the most computationally expensive step is the $\bw^{(t+1)}$ update that involves matrix inversion. This matrix inversion, however, can be computed efficiently using the %Sherman-Morrison-Woodbury 
identity $(\bI + \bA^{T}\bA)^{-1} = \bI - \bA^{T}(\bI + \bA\bA^{T})^{-1}\bA$. Since matrix $\bA$ is a fat matrix, $\bA\bA^{T}$ has smaller dimensions than $\bA^{T}\bA$. Moreover, one can easily see that $\bA\bA^{T}$ is a matrix of dimensions $(n+1) \times (n+1)$, and 
%\begin{align}
%\bA\bA^{T} &= 
%\begin{bmatrix}
%c_{n}		& 1			& 1			& \dots  & 1 \\
%1		& 1			& 0			& \dots  & 0 \\
%1		& 0			& 1 		& \dots  & 0 \\
%\vdots 	& \vdots 	& \vdots 	& \ddots & \vdots \\
%1		& 0			& 0			& \dots  & 1
%\end{bmatrix}.
%\end{align}
\begin{align}
\bA\bA^{T} &= 
\begin{bmatrix}
c_{n}		& \bone^{T}\\
\bone		& \bI
\end{bmatrix}.
\end{align}
where $c_{n} = 2n^2 - n$. In addition, the inverse only needs to be computed once at the start of the algorithm since this matrix is deterministic and depends only on the size of the adjacency matrix being estimated.

\subsection{Parameter and computational complexities}
The number of parameters that one needs to learn a graph adjacency matrix is $\frac{n(n+1)}{2}$. This implies that the number of unknown parameters scales quadratically with the number of nodes in the graph. Additionally, the per-iteration computational complexity of the proposed method also scales quadratically with the number of nodes \cite{wang2017new}. The same computational and memory complexities also hold for the existing state-of-the-art graph learning algorithms \cite{dong2016learning,kalofolias2017large,kalofolias2016learn,chepuri2017learning,egilmez2017graph}. However, while these complexities are manageable for small graphs, for real-world datasets with even hundreds of nodes current methods become prohibitive. To overcome these issues, we next examine the problem of learning product graphs.
%This implies that for learning unstructured graphs the computational complexity scales as $\rmO(n^{2}) = \rmO(\prod_{k=1}^{K_{0}} n_{k}^{2}) \sim \rmO(\bar{n}^{2K_{0}})$, assuming the special case of $n_{1} = n_{2} = \dots = n_{K_{0}} = \bar{n}$. On the other hand, when the product structure is imposed, one only has to solve $K_{0}$ smaller problems each with computational complexity of $\rmO(\bar{n}^{2})$. These computational gains are huge when compared to the original problem for learning unstructured graphs!

\section{Product graphs: Problem formulation \\\quad\& advantages in data representation} \label{sec:product_graphs}
In this section we briefly review product graphs and their implications towards graph learning. We investigate how product graphs provide a way to efficiently represent graphs with a huge number of nodes, and we revisit the notion of smoothness of signals over product graphs.

Let us consider $K_{0}$ graphs $G_{k} = \{V_{k}, E_{k}\}$, for $k = 1,\dots,K_{0}$, where $V_{k}$ and $E_{k}$ represent the vertices and edges of the $k$-th graph. The product of these graphs would result in a product graph $G = \{V, E\}$, with $V$ %$V \in \R^{n_{1} n_{2} \dots n_{K_{0}}}$
 and $E$ representing the vertices and edges of the resultant graph. The three most commonly investigated graph products and their respective adjacency matrices are discussed below. Note that graph adjacency matrices are considered in this work because each kind of product structure is directly reflected in adjacency matrix of the resultant graph.

\subsection{Kronecker graphs} \label{sec:product_graph_prob:kron}
For the Kronecker product of graphs $G_{k}$, for $k = 1,\dots,K_{0}$, with adjacency matrices $\bW_{k}$, the Kronecker product graph can be expressed as $G = \underset{[K_{0}]}{\otimes} G_{k} = G_{K_0} \otimes G_{K_0-1} \otimes \dots \otimes G_{1}$. The respective Kronecker-structured adjaceny matrix of the resultant graph can written in terms of component/factor adjacency matices as $\bW =\underset{[K_{0}]}{\otimes} \bW_{k}$. Additionally, if the factor adjacency matrix $\bW_{k}$ can be expressed via eigenvalue decomposition (EVD) as $\bW_{k} = \bU_{k} \bLambda_{k} \bU_{k}^{T}$, then the Kronecker adjaceny matrix can be written as (using properties in \cite{sandryhaila2014big}):
\begin{align} \label{eq:kronecker}
\bW	  &= (\bU_{K_{0}} \bLambda_{K_{0}} \bU_{K_{0}}^{T}) \otimes \dots \otimes (\bU_{1} \bLambda_{1} \bU_{1}^{T}) \nonumber\\
&= (\underset{[K_{0}]}{\otimes} \bU_{k})~(\underset{[K_{0}]}{\otimes} \bLambda_{k})~(\underset{[K_{0}]}{\otimes} \bW_{k}^{T}) = \bU \bLambda_{\mathrm{kron}} \bU^{T}.
\end{align}
One can see that both the eigenmatrix and the eigenvalue matrix of the Kronecker adjacency matrix have a Kronecker structure in terms of the component eigenmatrices and component eigenvalue matrices, respectively. Given the number of edges in the component graphs are $|E_{k}|$, the number of edges in the Kronecker graph are $|E| = 2^{K_{0}-1} \overset{K_{0}}{\underset{k = 1}{\prod}} |E_{i}|$. %A pictorial representation of edges present in the Kronecker product of graphs is shown (for $K_{0} = 2$) in Fig.~\ref{fig:product_graph}. 

An example of Kronecker product graph is the bipartite graph of a recommendation system like Netflix \cite{allen2010transposable} where the graph between users and movies can be seen as a Kronecker product of two smaller factor graphs. %The user graph in this case would capture the similarity among the user, while the graph of the movies will capture the similarities between the movies in the database. 
In fact, the adjacency matrix of any bipartite graph can be represented in terms of a Kronecker product of appropriate factor matrices \cite{leskovec2010kronecker}. As adjacency matrices are also closely related to precision matrices (inverse covariance matrices), and inverse of Kronecker product is Kronecker product of inverses \cite{sandryhaila2014big}, imposing Kronecker structure on the adjacency matrix also amounts to imposing a Kronecker structure on the covariance matrix of the data.

The optimization problem in \eqref{learn_lap_3} can be specialized to the case of learning Kronecker graphs by explicitly imposing the Kronecker product structure on the adjacency matrix and posing the problems in terms of the individual factor adjacency matrices, rather than the bigger adjacency matrix produced after the product. This leads us to the following nonconvex problem for learning Kronecker graphs:
\begin{align} \label{learn_kronecker}
\underset{\{\bW_{k} \in \cW \}_{k=1}^{K_{0}}}{\min} &\frac{\alpha}{M_{0}}\tr \Bigg[ \big[ \underset{[K_{0}]}{\otimes} \bW_{k} \big] \Big(\underset{m = 1}{\overset{M_{0}}{\sum}} \bone \bar{\bx}_{m}^{T} - \underset{m = 1}{\overset{M_{0}}{\sum}} \bx_{m} \bx_{m}^{T} \Big) \Bigg]. %\nonumber\\
%%\underset{\{\bW_{k}\}_{k=1}^{K_{0}}}{\min} &\frac{\alpha}{M_{0}}~\underset{m = 1}{\overset{M_{0}}{\sum}} \bar{\bx}_{m}^{T} (\underset{[K_{0}]}{\otimes} \bW_{k}) \bone - \bx_{m}^{T} (\underset{[K_{0}]}{\otimes} \bW_{k}) \bx_{m} \nonumber\\
%\text{s.t.}\quad 	&\diag(\bW_{k}) = \bzero,~ \bone^{T} \bW_{k} \bone = n_{k}, \nonumber\\
%&\quad\quad\quad\quad (\bW_{k})_{ij} = (\bW_{k})_{ji} \geq 0.
\end{align}

\subsection{Cartesian graphs} \label{sec:product_graph_prob:cart}
The Cartesian product (also called Kronecker sum product) of graphs $G_{k}$ is represented as $G = \underset{[K_{0}]}{\oplus} G_{k} = G_{K_0} \oplus G_{K_0-1} \oplus \dots \oplus G_{1}$. The correspoding cartesian adjacency matrix can be written in terms of the component adjacency matrices as $\bW =\underset{[K_{0}]}{\oplus} \bW_{k}$. Furthermore, with the EVD of the component adjacency matrices, the Cartesian adjacency matrix can be decomposed as \cite{sandryhaila2014big}:
\begin{align} \label{eq:cartesian}
\bW	  &= (\bU_{K_{0}} \bLambda_{K_{0}} \bU_{K_{0}}^{T}) \oplus \dots \oplus (\bU_{1} \bLambda_{1} \bU_{1}^{T}) \nonumber\\
&= (\underset{[K_{0}]}{\otimes} \bU_{k})~(\underset{[K_{0}]}{\oplus} \bLambda_{i})~(\underset{[K_{0}]}{\otimes} \bW_{k}^{T}) = \bU \bLambda_{\mathrm{cart}} \bU^{T}.
\end{align}
This means that the eigenmatrix and the eigenvalue matrix of the Cartesian adjacency matrix are represented, respectively, as Kronecker and Cartesian products of component eigenmatrices and eigenvalue matrices. The number of edges in the Cartesian graph can be found as $|E| = \sum_{k = 1}^{K_{0}} (\underset{[K_{0}] \setminus k}{\otimes} n_{i}) |E_{k}|$, where $|E_{k}|$ represents the number of edges and $n_{k}$ represents the number of vertices in the $k$-th component graph. %Fig.~\ref{fig:product_graph} shows the edges that are formed in as a result of Cartesian product of two graphs. 

A typical exmaple of a Cartesian product graph is images. Images reside on two-dimensional rectangular grids that can be represented as the Cartesian product between two line graphs pertaining to the rows and columns of the image \cite{sandryhaila2014big}. A social network can also be approximated as a Cartesian product of an inter-community graph with an intra-community graph \cite{sandryhaila2014big}.

Similar to the previous discussion, the optimization problem in \eqref{learn_lap_3} can be specialized to learning Cartesian graphs by explicitly imposing the Cartesian structure and posing the problem in terms of the factor adjacency matrices as follows:
\begin{align} \label{learn_cartesian}
\underset{\{\bW_{k} \in \cW \}_{k=1}^{K_{0}}}{\min} &\frac{\alpha}{M_{0}}\tr \Bigg[ \big[ \underset{[K_{0}]}{\oplus} \bW_{k} \big] \Big(\underset{m = 1}{\overset{M_{0}}{\sum}} \bone \bar{\bx}_{m}^{T} - \underset{m = 1}{\overset{M_{0}}{\sum}} \bx_{m} \bx_{m}^{T} \Big) \Bigg]. %\nonumber\\
%%\underset{\{\bW_{k}\}_{k=1}^{K_{0}}}{\min} &\frac{\alpha}{M_{0}}~\underset{m = 1}{\overset{M_{0}}{\sum}} \bar{\bx}_{m}^{T} (\underset{[K_{0}]}{\otimes} \bW_{k}) \bone - \bx_{m}^{T} (\underset{[K_{0}]}{\otimes} \bW_{k}) \bx_{m} \nonumber\\
%\text{s.t.}\quad 	&\diag(\bW_{k}) = \bzero,~ \bone^{T} \bW_{k} \bone = n_{k}, \nonumber\\
%&\quad\quad\quad\quad (\bW_{k})_{ij} = (\bW_{k})_{ji} \geq 0.
\end{align}
%Since the Cartesian product is linear in terms of factor adjacency matrices (see \eqref{eq:smoothness:cartesian}), the problem posed in \eqref{learn_cartesian} is actually convex.

\subsection{Strong graphs} \label{sec:product_graph_prob:strong}
The strong product of graphs $G_{k}$ can be represented as $G = \underset{[K_{0}]}{\boxtimes} G_{k} = G_{K_0} \boxtimes G_{K_0-1} \boxtimes \dots \boxtimes G_{1}$. The respective strong adjacency matrix of the resultant strong graph is given in terms of the component adjacency matrices as $\bW =\underset{[K_{0}]}{\boxtimes} \bW_{k}$, and can be further expressed as:
\begin{align} \label{eq:strong}
\bW	  &= (\bU_{K_{0}} \bLambda_{K_{0}} \bU_{K_{0}}^{T}) \boxtimes \dots \boxtimes (\bU_{1} \bLambda_{1} \bU_{1}^{T}) \nonumber\\
&= (\underset{[K_{0}]}{\otimes} \bU_{k})~(\underset{[K_{0}]}{\boxtimes} \bLambda_{i})~(\underset{[K_{0}]}{\otimes} \bW_{k}^{T}) = \bU \bLambda_{\mathrm{str}} \bU^{T},
\end{align}
in terms of EVD of the component adjacency matrices.

The strong product graphs can be seen as the sum of Kronecker and Cartesian products of the factor adjacency matrices. %Thus, the strong product has edges from both Kronecker and Cartesian products. A representation of the edges in a strong product of two graphs are shown in Fig.~\ref{fig:product_graph}. 
An example of data conforming to the strong product graph is a spatiotemporal sensor network graph, which consists of a strong product of a spatial graph and a temporal graph (representing the temporal dependencies of the sensors). The spatial graph has as many nodes as the number of sensors in the sensor network and represents the spatial distribution of sensors. On the other hand, the temporal graph has as many nodes as the number of temporal observations of the whole sensor network and represents the overall temporal dynamics (changes in connectivity over time) of the network \cite{sandryhaila2014big}.

By making the strong product structure explicit in terms of the factor adjacency matrices, the optimization problem for learning strong graphs can be expressed as the following nonconvex problem:
\begin{align} \label{learn_strong}
\underset{\{\bW_{k} \in \cW \}_{k=1}^{K_{0}}}{\min} &\frac{\alpha}{M_{0}}\tr \Bigg[ \big[ \underset{[K_{0}]}{\boxtimes} \bW_{k} \big] \Big(\underset{m = 1}{\overset{M_{0}}{\sum}} \bone \bar{\bx}_{m}^{T} - \underset{m = 1}{\overset{M_{0}}{\sum}} \bx_{m} \bx_{m}^{T} \Big) \Bigg]. %\nonumber\\
%%\underset{\{\bW_{k}\}_{k=1}^{K_{0}}}{\min} &\frac{\alpha}{M_{0}}~\underset{m = 1}{\overset{M_{0}}{\sum}} \bar{\bx}_{m}^{T} (\underset{[K_{0}]}{\otimes} \bW_{k}) \bone - \bx_{m}^{T} (\underset{[K_{0}]}{\otimes} \bW_{k}) \bx_{m} \nonumber\\
%\text{s.t.}\quad 	&\diag(\bW_{k}) = \bzero,~ \bone^{T} \bW_{k} \bone = n_{k}, \nonumber\\
%&\quad\quad\quad\quad (\bW_{k})_{ij} = (\bW_{k})_{ji} \geq 0.
\end{align}

\subsection{Product  graph Fourier transform}
One can see from \eqref{eq:kronecker}, \eqref{eq:cartesian}, and \eqref{eq:strong} that the graph Fourier transform of a product graph (which is the eigenmatrix of the product adjacency matrix), has a Kronecker structure in terms of the eigenmatrices of the component graph adjacency matrices: $\bU = \underset{[K_{0}]}{\otimes} \bU_{k}$. In terms of the implementation of the graph Fourier transform, this structure provides an efficient implementation of the graph Fourier as (using the properties of Kronecker product and tensors \cite{kolda2009tensor}):
\begin{align}
\bU^{T} \bx = (\underset{[K_{0}]}{\otimes} \bU_{k})^{T} \bx = \tvec (\cX \underset{[K_{0}]}{\times} \bU_{k}),
\end{align}
where $\bx \in \R^{n_{1} n_{2} \dots n_{K_{0}}}$ is an arbitrary graph signal on the product graph, and $\cX \in \R^{n_{1} \times n_{2} \times n_{K_{0}}}$ represents appropriately tensorized version of the signal $\bx$. Because of this, one does not need to form the huge Fourier matrix $\bU$ and can avoid costly matrix multiplications by just applying the component graph Fourier matrices to each respective mode of the tensorized observation $\cX$ and then vectorizing the result.

\subsection{Smoothness } \label{sec:product_graphs:smoothness}
Smoothness of a graph signal is one of the core concepts in graph signal processing \cite{bronstein2017geometric,sandryhaila2013discrete,sandryhaila2014big,shuman2013emerging,ortega2018graph} and product graph Laplacians provide an efficient representation for the notion of smoothness. The smoothness of a graph signal can be measured through the Dirchlet energy defined as $\bx^{T} \bL \bx$. The Dirichlet energy can be reexpressed as: $\bx^{T} \bL \bx = \bx^{T} (\bD - \bW) \bx = \bx^{T} \bD \bx - \bx^{T} \bW \bx.$
Let us now focus on each term separately in the context of product graphs. For the term involving $\bW$ we have:
\begin{align} \label{eq:product_graphs:dirchilet:adjacency}
\bx^{T} \bW \bx	   &= \bx^{T} \bU \bLambda \bU^{T} \bx = (\bU^{T}\bx)^{T} \bLambda (\bU^{T}\bx) \nonumber\\
&= \tvec (\cX \underset{[K_{0}]}{\times} \bU_{k})^{T} \bLambda \tvec (\cX \underset{[K_{0}]}{\times} \bU_{k}).
\end{align} \label{eq:product_graphs:dirchilet:degree}
Similarly, the term involving $\bD$ can be re-expressed as:
\begin{align}
\bx^{T} \bD \bx		&= \bx^{T} \diag( \bW \bone) \bx = \bx^{T} \diag(\bx) \bW \bone \nonumber\\
&= (\bx \odot \bx)^{T} \bW \bone = \bar{\bx}^{T} \bW \bone, %\nonumber\\
%&= \tvec (\bar{\cX} \underset{[K_{0}]}{\times} \bU_{k})^{T} \bLambda \tvec (\mathds{1} \underset{[K_{0}]}{\times} \bU_{k}),
\end{align}
which can be computed efficiently along the lines of \eqref{eq:product_graphs:dirchilet:adjacency}. With this reformulation, one circumvents the need to explicitly form the prohibitively large eigenmatrix $\bU$ and can evaluate the Dirichlet energy much more efficiently with just mode-wise products with the smaller component eigenmatrices.

\subsection{Representation complexity} \label{sec:product_graphs:representation}
Let us consider an unknown graph $G$  with the number of nodes $|V| = n = \prod_{k =1}^{K_{0}} n_{k}$, where $n_{k}$ represents the number of nodes in each component graph and $K_{0}$ is total number of component graphs. If one were to learn this graph by means of an arbitrary adjacency matrix, the number of parameters that needs to be estimated would be $\frac{n(n + 1)}{2}$ (since the graph adjacency matrix is a symmetric matrix). On the other hand, for the same graph, by utilizing the product model of the graph adjacency matrix, one would need to estimate only $\sum_{k=1}^{K_{0}} \frac{n_{k}(n_{k} + 1)}{2}$ parameters. Considering the special case of $n_{1} = n_{2} = \cdots = n_{K_{0}}= \bar{n}$, and imposing the product structure on graph adjacency matrix reduces the number of parameters needed to be learned by $\bar{n}^{K_{0}-1}/K_{0}$!

%% file: learning_algo.tex
\begin{algorithm}[h]
	\caption{\textbf{: B-PGL---BCD for product graph learning}}
	\label{alg_learn_product_graph}
	{\textbf{Input:}  Observations $\{\bx_{m}\}_{m = 1}^{M_{0}} $, maximum iterations $N_{0}$, maximum inner iterations $T_{0}$, and parameters $\alpha$, $\rho$}\\
	\textbf{Initialize:} $\{\whbW_{k}\}_{k=1}^{K_{0}}$
	\begin{algorithmic}
		\STATE \textbf{for} $n=1$ to $N_{0}$ (outer interations)
		\STATE \hspace{\algorithmicindent}\textbf{for} $k=1$ to $K_{0}$
		\STATE \hspace{\algorithmicindent}\hspace{\algorithmicindent}\textbf{If} stopping criteria \textbf{not met}
		\STATE \hspace{\algorithmicindent}\hspace{\algorithmicindent} Solve \eqref{learn_kronecker}, \eqref{learn_cartesian}, or \eqref{learn_strong} for $\whbW_{k}$ via Algorithm~\ref{alg_learn_graph} %with $T_{0}$ inner iterations and parameters $\alpha$, $\rho$
		\STATE \hspace{\algorithmicindent}\hspace{\algorithmicindent}\textbf{end}
		\STATE \hspace{\algorithmicindent}\textbf{end}
		\STATE \textbf{end}
	\end{algorithmic}
	\textbf{Output:} Final adjacency matrix estimates $\whbW_{k}$
\end{algorithm}

\section{Algorithm for learning product graphs} \label{sec:learning_algo}
%Through the preceding subsections, we can see that solving factorwise minimization problems leads one closer to the true solutions of the factor adjacency matrices, up to some error. However, there were no restrictive initialization assumptions on the factor adjacency matrices that were fixed for each factorwise problems. Even so, starting with a better initialization produces a lower error. Thus to get closer to the true solution, we need to optimize over all factor matrices several times, each time with a better estimate. This leads us to our \textbf{B}lock coordinate descent algorithm for \textbf{P}roduct \textbf{G}raph \textbf{L}earning (B-PGL), outlined in Algorithm~\ref{alg_learn_product_graph}.

In the previous section we highlighted some properties and advantages of product graphs and we posed the optimization problems for learning these graphs. We now propose an algorithm for solving these product graph learning problems. To this end, we first recognize that even though the problems posed in \eqref{learn_kronecker}, \eqref{learn_cartesian}, and \eqref{learn_strong} are nonconvex, minimization of each objective with respect to any single factor adjacency matrix is still convex (while all the other factors matrices are fixed). Moreover, these factor-wise minimization problems can be solved through Algorithm~\ref{alg_learn_graph} proposed in the earlier sections. These observations lead us to propose a block coordinate descent (BCD) based algorithm, named B-PGL (BCD for product graph learning), that minimizes over each factor adjacency matrix in cyclic fashion. The proposed algorithm is provided in Algorithm~\ref{alg_learn_product_graph}, and in the following discussion we present the factor-wise problems for each product graph.

\subsection{Kronecker graphs}
Since Algorithm~\ref{alg_learn_product_graph} utilizes factor-wise minimization, we can characterize the error for product graph learning in terms of the factor-wise errors of each factor adjacency matrix. %(while keeping the other factors fixed). The factor-wise minimization problem in the case of learning Kronecker graphs boils down to (see Appendix~\ref{app:th:kronecker}):
%\begin{align} \label{learn_kronecker_factor}
%\underset{\bW_{k} \in \cW }{\min} &~\alpha~\tr(\bD_{k} \bar{\bS}_{k}) - \alpha~\tr\big( \bW_{k} \bS_{k} \big) %\nonumber\\
%%\text{s.t. } 	&\diag(\bW_{k}) = \bzero, \bone^{T} \bW_{k} \bone = n_{k},(\bW_{k})_{ij} = (\bW_{k})_{ji} \geq 0,
%\end{align}
%for $k = 1,\cdots,K_{0}$, and where $\bS_{k}$ and $\bar{\bS}_{k}$ are as defined in Appendix~\ref{app:th:kronecker}. 
%As pointed out before, each of these factor-wise problem is a convex program. 
The error bounds for learning Kronecker graphs are provided in the following theorem with the proof in  Appendix~\ref{app:th:kronecker}.
%\begin{theorem} \label{th:kronecker}
%For the estimate $\whbW_{k}$ of the ${k}$-th factor adjacency matrix comprising a Kronecker product adjacency matrix, obtained after each outer iteration in Algorithm~\ref{alg_learn_product_graph}, %while using the estimates $\whbW_{j}$ for $j = 1,\cdots,K_{0}, j \neq k$ of the other components, 
%the error between the sample-based minimization (with $M_{0}$ samples) and the population-based minimization (with infinitely many samples) of \eqref{learn_kronecker} satisfies $\cO \Big(\frac{n_{k}^{2}\log(n_{k})}{nM_{0}} \Big(1+\big\|\underset{[K_{0}] \setminus k}{\otimes} \whbW_{i} - \underset{[K_{0}] \setminus k}{\otimes} \bW_{i} \big\|_{F} \Big) \Big)$ with high probability as $\frac{nM}{n_{k}} \rightarrow \infty$ for $\alpha \leq \sqrt{\frac{n_{k}\log(n_{k})}{nM_{0}}}$.
%%\begin{align*}
%%\Big|\tr\big( \bL_{k} \bS_{k} \big) - \E \big[ \tr\big( \bL_{k} \bS_{k} \big) \big]\Big| 
%%= \cO_{P} \Bigg(\sqrt{\frac{n_{k}^{3}\log(n_{k})}{nM_{0}}} \Bigg).
%%\end{align*}
%Moreover, the error between the estimated factor $\whbW_{k}$ and the true factor $\bW_{k}$ satisfies $\| \whbW_{k} - \bW_{k}\|_{F} = \cO \Bigg( \sqrt{\frac{n_{k}\log(n_{k})}{nM_{0}}} \Bigg)$.
%%\begin{align} \label{th:kronecker:eq}
%%\| \whbW_{k} - \bW_{k}\|_{F} = \cO_{P} \Bigg( \sqrt{\frac{n_{k}\log(n_{k})}{nM_{0}}} \Bigg).
%%\end{align}
%\end{theorem}
\begin{theorem} \label{th:kronecker}
After each outer iteration of Algorithm~\ref{alg_learn_product_graph}, the error between the sample-based minimization (with $M_{0}$ samples) and the population-based minimization (with infinitely many samples) of \eqref{learn_kronecker} with respect to the ${k}$-th factor $\bW_{k}$ satisfies $\cO \Big(\frac{n_{k}^{2}\log(n_{k})}{nM_{0}} \Big(1+\big\|\underset{[K_{0}] \setminus k}{\otimes} \whbW_{i} - \underset{[K_{0}] \setminus k}{\otimes} \bW_{i}^{*} \big\|_{F} \Big) \Big)$, with high probability as $\frac{nM}{n_{k}} \rightarrow \infty$, when $\alpha \leq \sqrt{\frac{n_{k}\log(n_{k})}{nM_{0}}}$.
Additionally, the error between the estimated factor $\whbW_{k}$ and the original generating factor $\bW_{k}^{*}$ also satisfies $\| \whbW_{k} - \bW_{k}^{*}\|_{F} = \cO \Bigg( \sqrt{\frac{n_{k}\log(n_{k})}{nM_{0}}} \Bigg)$, with high probability as $\frac{nM}{n_{k}} \rightarrow \infty$.
\end{theorem}
\begin{remark}
The error between the estimated $k$-th factor and the original generating $k$-th factor for learning Kronecker graphs also depends on the estimation errors of the other factors. This dependence, however, is absorbed in big-$\cO$.
\end{remark}

\subsection{Cartesian graphs}
%For Cartesian graphs, the factor-wise minimization problems, for $k = 1,\cdots,K_{0}$, can be represented as follows (see Appendix~\ref{app:th:cartesian}):
%\begin{align} \label{learn_cartesian_factor}
%\underset{\bW_{k} \in \cW }{\min} &~\alpha~\tr(\bD_{k} \bar{\bT}_{k}) - \alpha~\tr\big( \bW_{k} \bT_{k} \big) %\nonumber\\
%%\text{s.t. } 	&\diag(\bW_{k}) = \bzero, \bone^{T} \bW_{k} \bone = n_{k},(\bW_{k})_{ij} = (\bW_{k})_{ji} \geq 0,
%\end{align}
%where $\bT_{k}$ and $\bar{\bT}_{k}$ are as defined in Appendix~\ref{app:th:cartesian}. As before, each factor-wise problem is a convex program, and 
The following theorem characterizes the error of the factor-wise minimization of the Cartesian graph learning problem.
\begin{theorem} \label{th:cartesian}
The objective function \eqref{learn_cartesian} for the Cartesian graph learning problem can be represented as a sum of terms that are linear in each factor adjacency matrix, and is therefore convex. 
After each outer iteration of Algorithm~\ref{alg_learn_product_graph}, the error between the sample-based minimization (with $M_{0}$ samples) and the population-based minimization (with infinitely many samples) of \eqref{learn_cartesian} with respect to the ${k}$-th factor $\bW_{k}$ satisfies $\cO \Big(\frac{n_{k}^{2}\log(n_{k})}{nM_{0}} \Big)$, with high probability as $\frac{nM}{n_{k}} \rightarrow \infty$, when $\alpha \leq \sqrt{\frac{n_{k}\log(n_{k})}{nM_{0}}}$.
Additionally, the error between the estimated factor $\whbW_{k}$ and the original generating factor $\bW_{k}^{*}$ also satisfies $\| \whbW_{k} - \bW_{k}^{*}\|_{F} = \cO \Bigg( \sqrt{\frac{n_{k}\log(n_{k})}{nM_{0}}} \Bigg)$, with high probability as $\frac{nM}{n_{k}} \rightarrow \infty$.
\end{theorem}
The proof of this theorem is given in Appendix~\ref{app:th:cartesian}. The theorem states, remarkably, that the objective function for learning Cartesian product graphs is convex and separable in each factor adjacency matrix, i.e., the objective function can be represented as a sum of linear terms, each of which is dependent on only one factor adjacency matrix. As a consequence of this fact, for learning Cartesian graphs, one can optimize over all factor adjacency matrices in parallel!
\begin{remark}
The error between the estimated $k$-th factor and the original generating $k$-th factor for learning Cartesian graphs is independent of the estimation errors of the other factors due to the convexity and separability of \eqref{learn_cartesian}.
\end{remark}
%The error for each factor Laplacian in Cartesian Laplacian can be characterized along the lines of Theorem~\ref{th:kronecker}, using matrix concentration inequalities (see \cite{sun2015non}).% based on the Hanson-Wright inequality \cite{greenewald2017tensor,rudelson2013hanson}.

\subsection{Strong graphs}
%The problem for learning strong graphs can be posed factor-wise, for $k = 1,\cdots,K_{0}$, (see Appendix~\ref{app:th:strong}) as follows:
%\begin{align} \label{learn_strong_factor}
%\underset{\bW_{k} \in \cW }{\min} &~\alpha~\tr(\bD_{k} \bar{\bZ}_{k}) - \alpha~\tr\big( \bW_{k} \bZ_{k} \big) %\nonumber\\
%%\text{s.t. } 	&\diag(\bW_{k}) = \bzero, \bone^{T} \bW_{k} \bone = n_{k},(\bW_{k})_{ij} = (\bW_{k})_{ji} \geq 0.
%\end{align}
%where $\bZ_{k}$ and $\bar{\bZ}_{k}$ are as defined in Appendix~\ref{app:th:strong}. 
The following theorem, with its proof in Appendix~\ref{app:th:strong}, characterizes the behavior of factor-wise minimization for learning strong product graphs.
\begin{theorem} \label{th:strong}
After each outer iteration of Algorithm~\ref{alg_learn_product_graph}, the error between the sample-based minimization (with $M_{0}$ samples) and the population-based minimization (with infinitely many samples) of \eqref{learn_strong} with respect to the ${k}$-th factor $\bW_{k}$ satisfies $\cO \Big(\frac{n_{k}\log(n)}{M_{0}} \Big({1+2\big\|\underset{[K_{0},k]}{\boxtimes} \whbW_{i} - \underset{[K_{0}]}{\boxtimes} \bW_{i}^{*} \big\|_{F}} \Big) \Big)$, with high probability as $\frac{nM}{n_{k}} \rightarrow \infty$, when $\alpha \leq \sqrt{\frac{n_{k}\log(n_{k})}{nM_{0}}}$.
Additionally, the error between the estimated factor $\whbW_{k}$ and the original generating factor $\bW_{k}^{*}$ also satisfies $\| \whbW_{k} - \bW_{k}^{*}\|_{F} = \cO \Bigg( \sqrt{\frac{n_{k}\log(n_{k})}{nM_{0}}} \Bigg)$, with high probability as $\frac{nM}{n_{k}} \rightarrow \infty$.
%
%For the ${k}$-th adjacency factor comprising a strong product adjacency matrix, while keeping other components $\bW_{j}$ for $j = 1,\cdots,K_{0}, j \neq k$ fixed, the error between the sample-based minimization with $M_{0}$ samples and the population-based minimization of \eqref{learn_strong_factor} satisfies $\cO_{P} \Big(\frac{n_{k}\log(n)}{M_{0}} \Big({1+2\big\|\underset{[K_{0},k]}{\boxtimes} \whbW_{i} - \underset{[K_{0}]}{\boxtimes} \bW_{i} \big\|_{F}} \Big) \Big)$, for an appropriately chosen $\alpha$.
%Moreover, the error between the estimated factor $\whbW_{k}$ and the true factor $\bW_{k}$ satisfies $\| \whbW_{k} - \bW_{k}\|_{F} = \cO_{P} \Bigg( \sqrt{\frac{n_{k}\log(n_{k})}{nM_{0}}} \Bigg)$.
\end{theorem}
\begin{remark}
The error between the estimated $k$-th factor and the original generating $k$-th factor for learning strong graphs is dependent on the estimation errors of the other factors. This dependence on other factors, however, is absorbed in big-$\cO$.
\end{remark}

%\begin{remark}
%Each theorem only provides the estimation error bounds of factor adjacency matrices for respective product graph. Error bounds on the product adjacency matrix are non-trivial and will be the focus of future works. However, intuitively speaking, the error for estimating Cartesian graphs should be smaller than other product graphs as the Cartesian adjacency matrix can be obtained as a linear combination of the factor matrices, whereas Kronecker and strong graphs contain terms obtained through products of the factor adjacency matrices.
%\end{remark}

\subsection{Error bound for arbitrary graphs} \label{sec:product_graph_prob:arbitrary}
As a byproduct of Theorem~\ref{th:kronecker}, we can also obtain an error bound for arbitrary graph learning problem. To this end, see that the objective in \eqref{learn_lap_3} for learning arbitrary graphs can be expressed as $\alpha/M_{0} \sum_{m=1}^{M_{0}}\bx_{m}^{T} \bL \bx_{m} = \alpha \tr( \bD \bar{\bS}) - \alpha \tr( \bW \bS)$, where $\bS$ and $\bar{\bS}$ are similiar to the definitions in Appendix~\ref{app:th:kronecker}. % = \alpha \tr( \bW \wtbS)$, through algebraic manipulation. %where the matrix $\bS = \frac{1}{M_{0}}\sum_{m = 1}^{M_0} \bx_{m} \bx_{m}^{T}$. 
%With this, the form of the unstructured graph learning problem looks similar to the factorwise problems for structured graphs in \eqref{learn_kronecker_factor}, \eqref{learn_cartesian_factor}, and \eqref{learn_strong_factor}. 
Following along the lines of Theorem~\ref{th:kronecker}, by solving \eqref{learn_lap_3} via Algorithm~\ref{alg_learn_graph} one is guaranteed to converge to the original generating adjacency matrix of the unstructured graph with the error $\| \whbW - \bW^{*}\|_{F} = \cO_{P} \Big( \sqrt{\log(n)/M_{0}} \Big)$, with a high probability as $M \rightarrow \infty$.

\subsection{Computational complexity}
The computational complexity of solving each factor-wise problem scales quadratically with the number of nodes in the graph. This implies that when the product structure is imposed, one only has to solve $K_{0}$ smaller problems each with computational complexity of $\rmO(\bar{n}^{2})$, assuming the special case of $n_{1} = n_{2} = \dots = n_{K_{0}} = \bar{n}$. In contrast, for learning unstructured graphs the computational complexity would scale as $\rmO(n^{2}) = \rmO(\prod_{k=1}^{K_{0}} n_{k}^{2}) \sim \rmO(\bar{n}^{2K_{0}})$. Thus, the computational gains are huge in comparison to the original problem for learning unstructured graphs!

This computational gain is only made possible due to the way that we pose the original graph learning problem as a linear program. This way of posing the original problem made the objective amenable to factor-wise minimization, which would not be possible if the original graph learning problem was posed in the traditional way from the existing literature.

\subsection{Convergence properties \& sample complexity}
%Each of the preceding theorems derive the error bounds after the first iteration of Algorithm~\ref{alg_learn_product_graph} for each product structure. 
The overall convergence of the algorithm to a stationary point of the problem can be established through the following theorem, whose proof is provided in Appendix~\ref{app:th:convergence}.
\begin{theorem} \label{th:convergence}
Since each factor-wise problem for each graph learning problem is convex, the product graph learning algorithm Algorithm~\ref{alg_learn_product_graph} is guaranteed to converge to a stationary point at a linear rate.
\end{theorem}
%The proof of this theorem is provided in Appendix~\ref{app:th:convergence}.

%\begin{remark}
We now provide an insight into our theorems statements with regards to the required number of observation. Theorem~\ref{th:kronecker}, \ref{th:cartesian}, and \ref{th:strong} claim that the estimated factor lies within a ball of radius $\frac{\log(n_{k})}{(n/n_{k})M_{0}}$ around the true factor. The accuracy of the estimate increases with the number of available observations $M_0$, and the product of the dimensions of the other factors $n/n_{k} = \prod_{j\neq k} n_{j}$. Moreover, the accuracy decreases with the increasing dimensions of the factor being estimated.
%\end{remark}

%\begin{remark}
Taking a closer look at the error bounds for learning arbitrary and structured graphs reveals an important point. The denominator in the error bound is the number of observations available to estimate the graph. For $M_{0}$ observed graph signals, the number of observations available to arbitrary graph learning are (obviously) $M_{0}$; however, for estimating the $k$-th factor adjacency when learning product graphs the effective number of observations are $\prod_{j\neq k} n_{j} \times M_{0}$. This means that imposing the product structure results in an increased number of effective observations to estimate each factor adjacency matrix. This combined with the reduced number of parameters required to learn these graphs makes product graphs very attractive for real world applications.
%By imposing the product structure on graph, each graph signal observation results in effectively $\prod_{j\neq k} n_{j}$ observations for estimating the $k$-th component adjacency matrix. In contrast, for learning an arbitrary adjacency matrix each observation only counts as one since we are trying to learn one large adjacency matrix instead of several smaller ones. This tradeoff of imposing the structure can be seen through our Theorem~\ref{th:kronecker} which states that the error for learning a factor graph scales as $\cO_{P} \Big( \sqrt{\frac{n_{k}\log(n_{k})}{nM_{0}}} \Big)$. The error for learning the unstructured adjacency matrix, however, scales as $\cO_{P} \Big( \sqrt{\frac{\log(n)}{M_{0}}} \Big)$ (which can also be inferred from Theorem~\ref{th:kronecker}). This line of reasoning and exploiting product structure has also been successfully explored in the dictionary learning literature; see \cite{ghassemi2019learning,shakeri2018identifiability} and the references therein for details.
%\end{remark}

%% file: numerical_experiments.tex
\section{Numerical experiments} \label{sec:numerical_experiments}
This section provides results for learning product graphs from synthetic and real datasets. We first present experiments for learning arbitrary graphs through our proposed linear program in Sec.~\ref{sec:graph_learning_linear}, and then the results for learning products graphs from synthetic data through Alg.~\ref{alg_learn_product_graph}. Afterwards we validate the performance of our proposed algorithm for product graphs on real-world datasets.
%%%%%%%%%%%%%%%%%%%%%%%%%%%%%%%%%%%%%%%%%%%%%%%%%%%%%%%%%%%%%%%%%%%%%%%%%%%%%%%%%%%%%%%%%%%%%%%%%%%%%%%%%%%%%%%%%%%%%%%%%%%%%%%%%%%%%%%%%%%%%%%%%%%%%%
\begin{figure*} [th]
	\begin{center}
		\begin{tabular}{cccc}
			{\includegraphics[height=2.6cm] {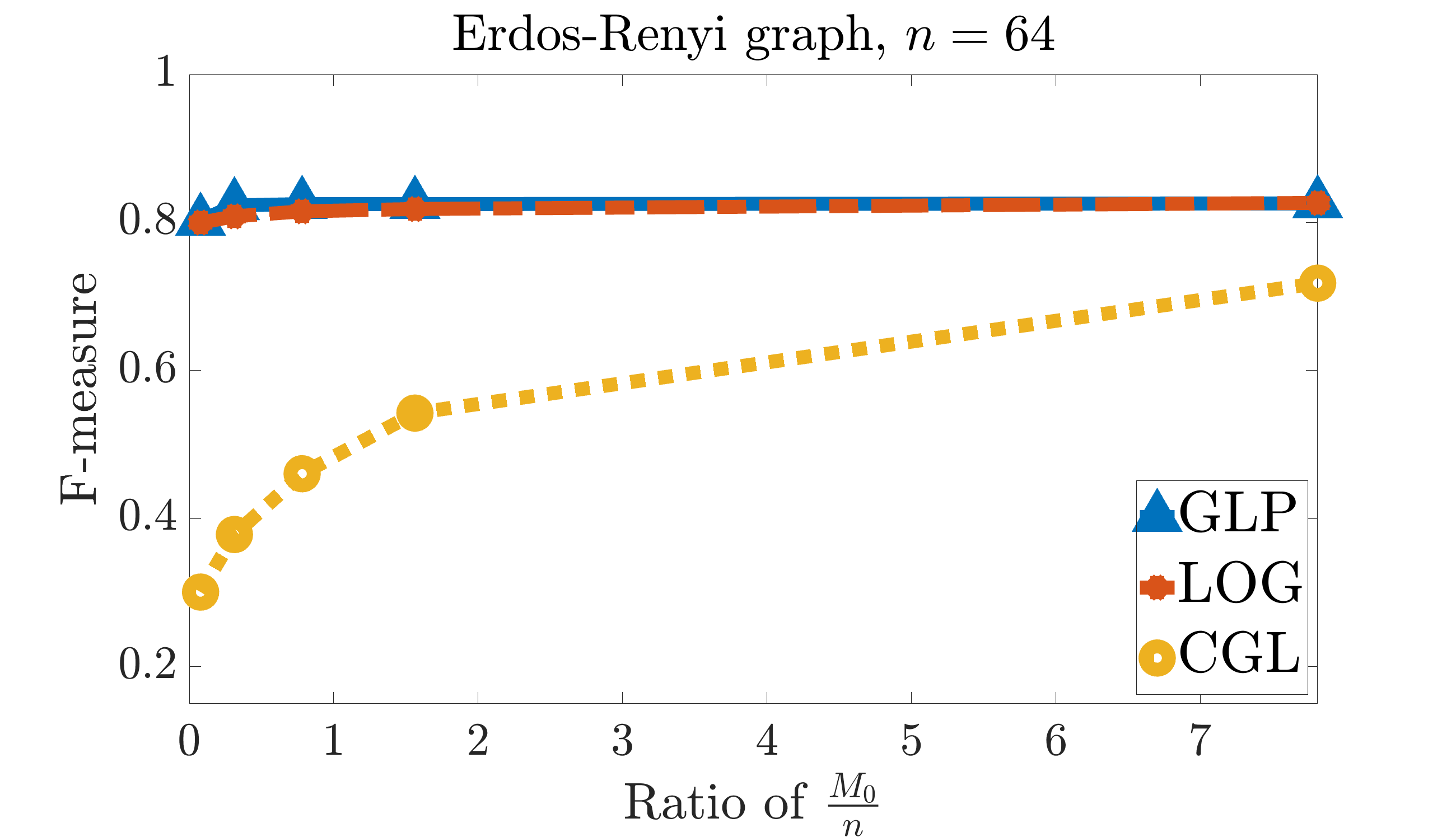}}
			{\includegraphics[height=2.6cm] {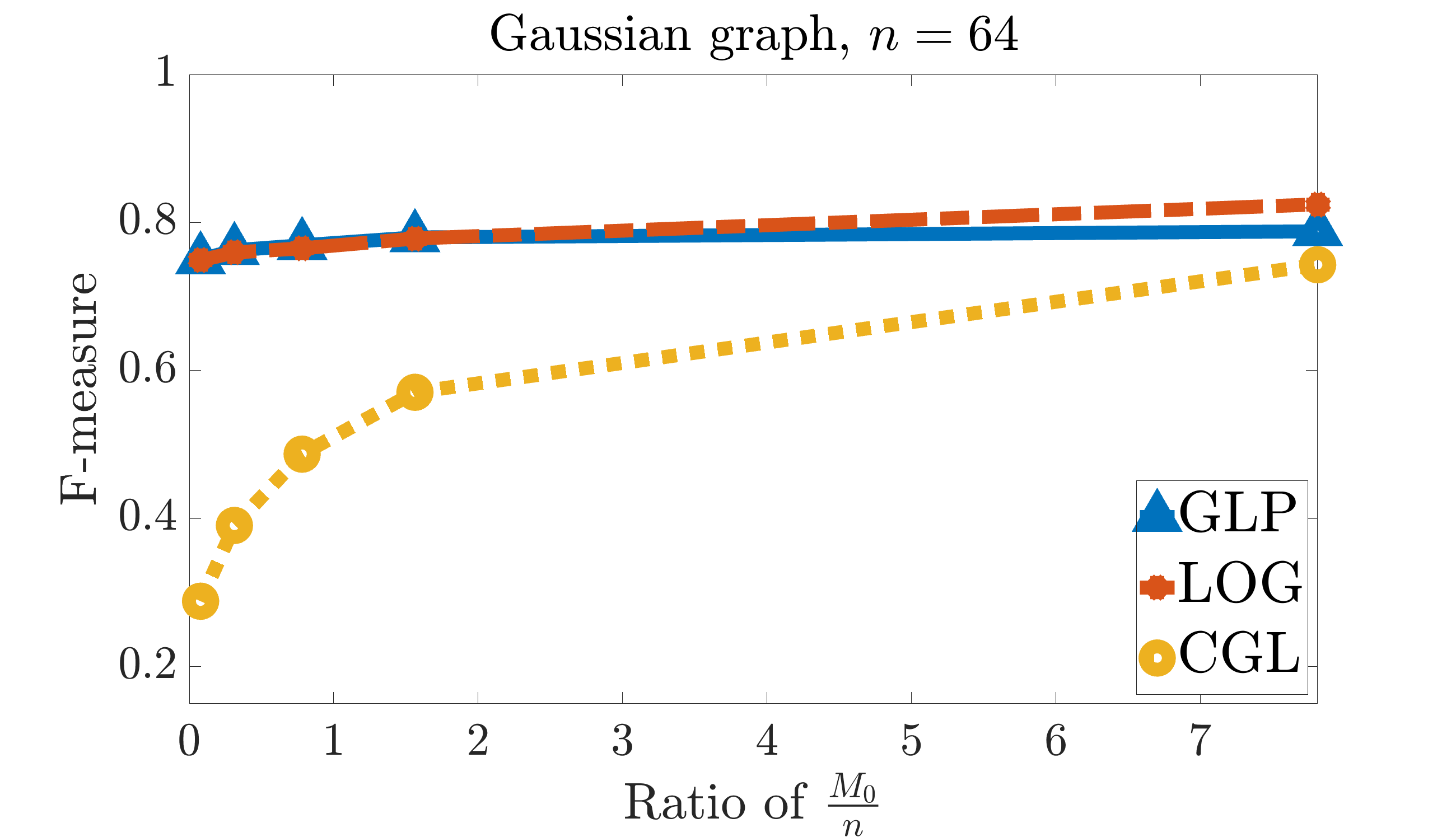}}
			{\includegraphics[height=2.6cm] {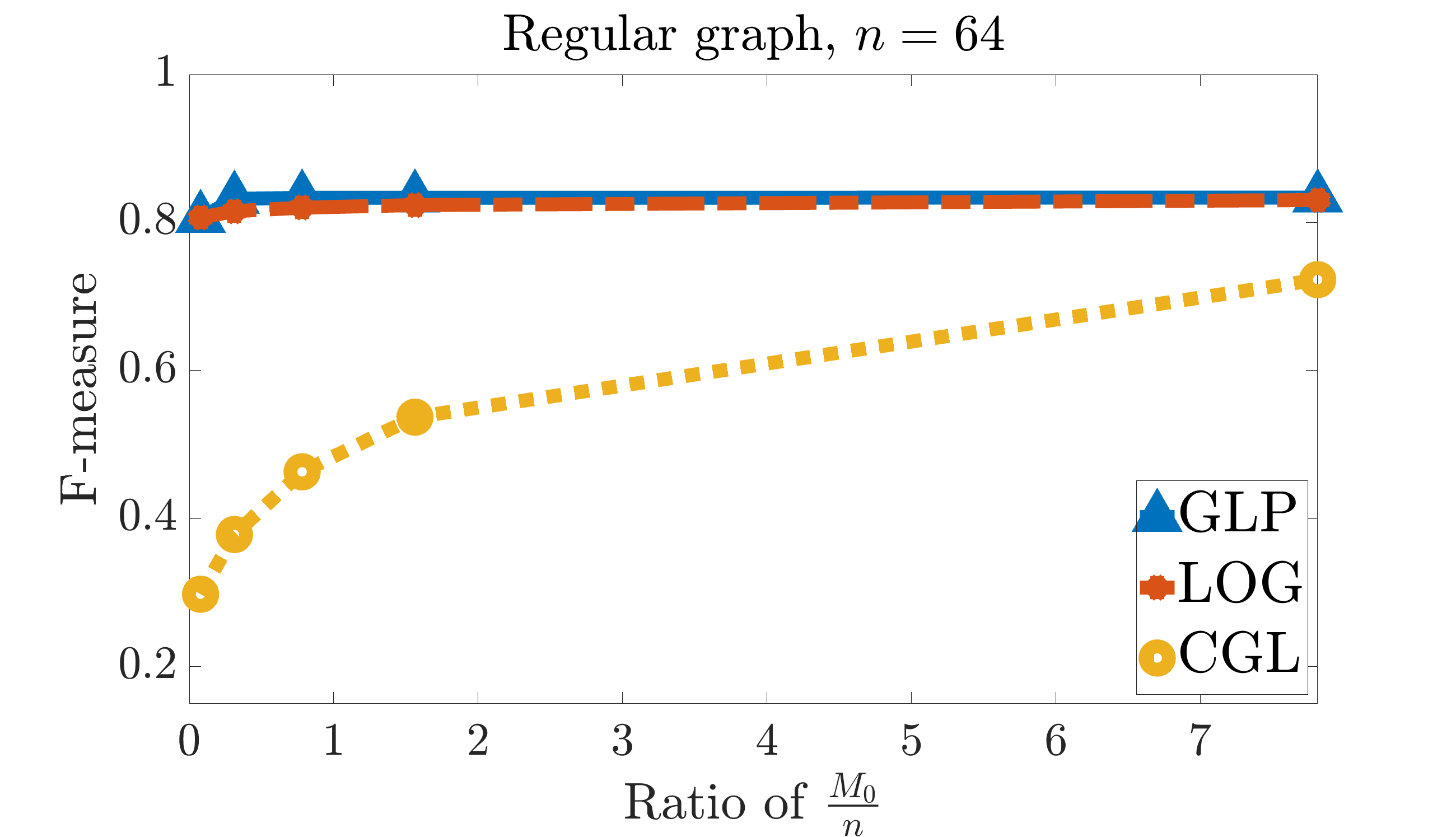}}
			{\includegraphics[height=2.6cm] {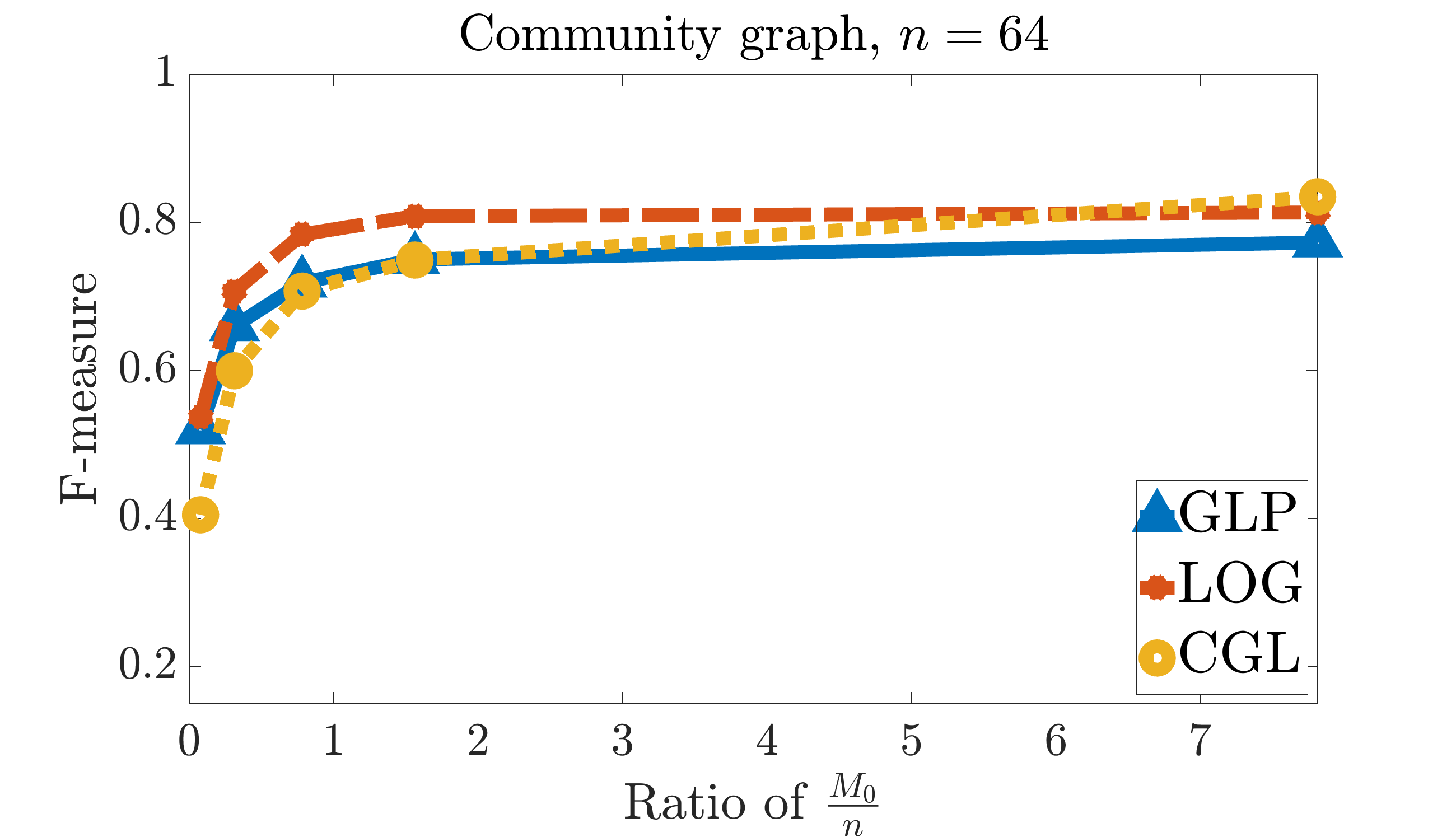}}\\
			{\includegraphics[height=2.6cm] {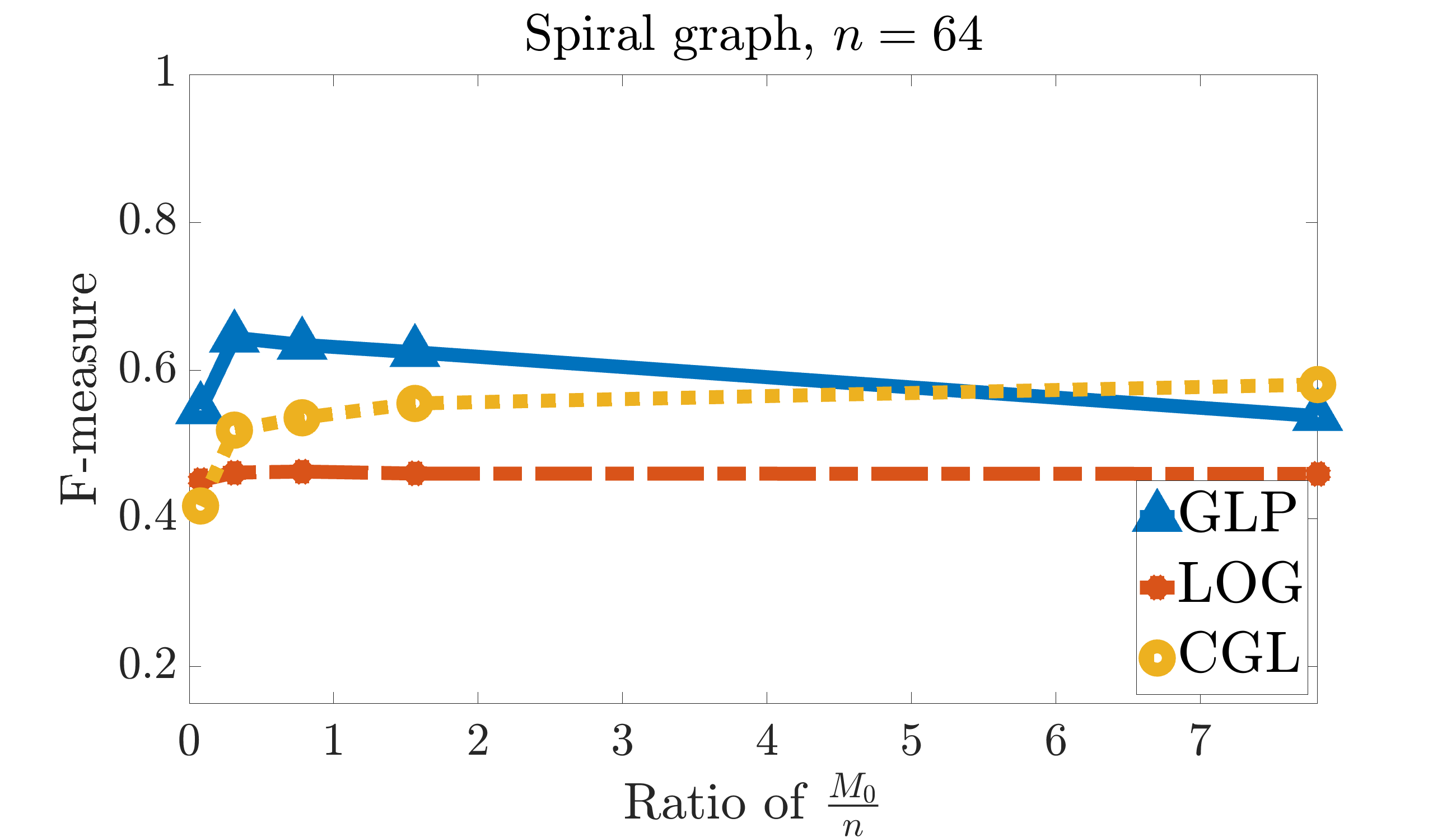}}
			{\includegraphics[height=2.6cm] {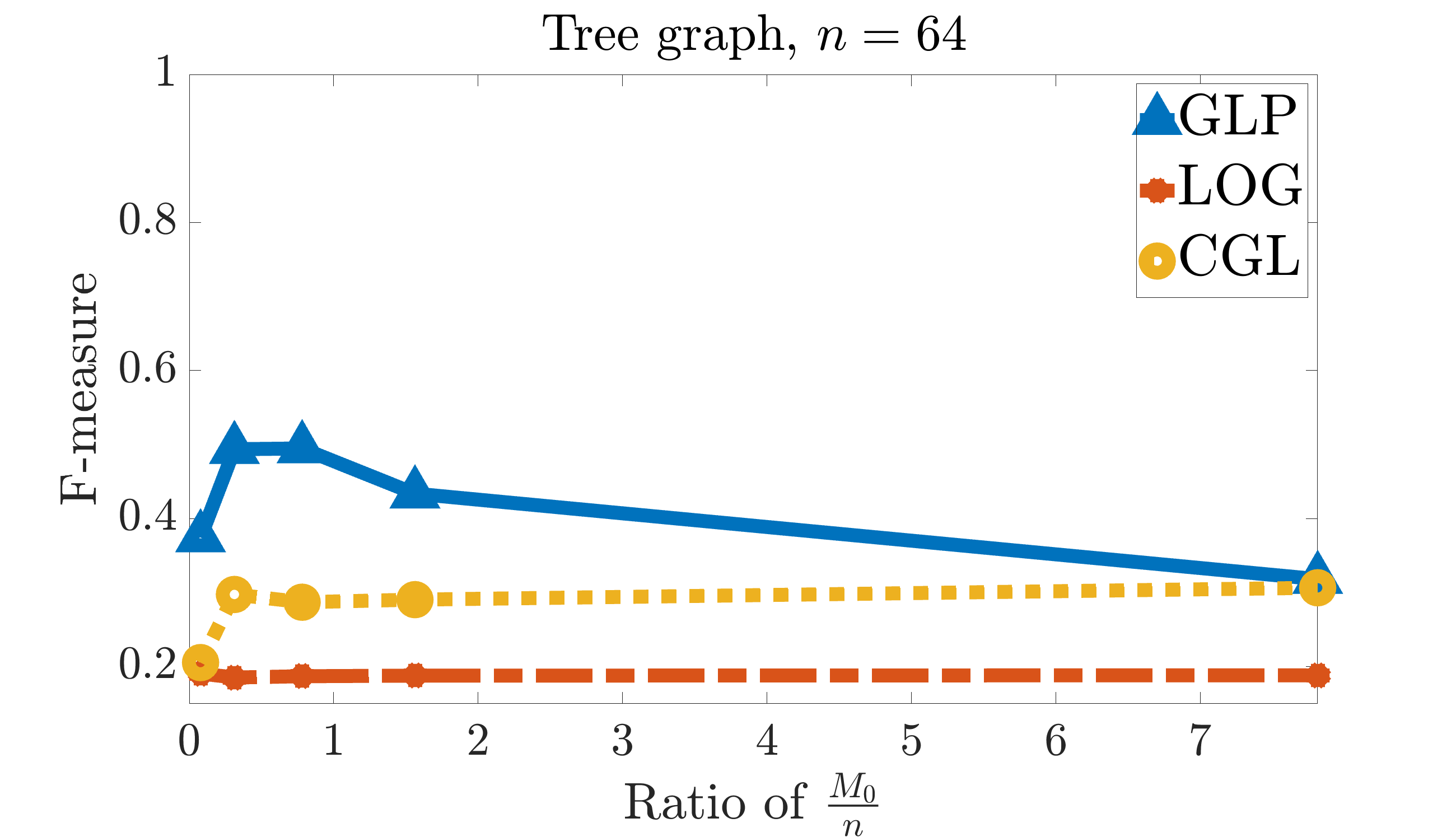}}
			{\includegraphics[height=2.6cm] {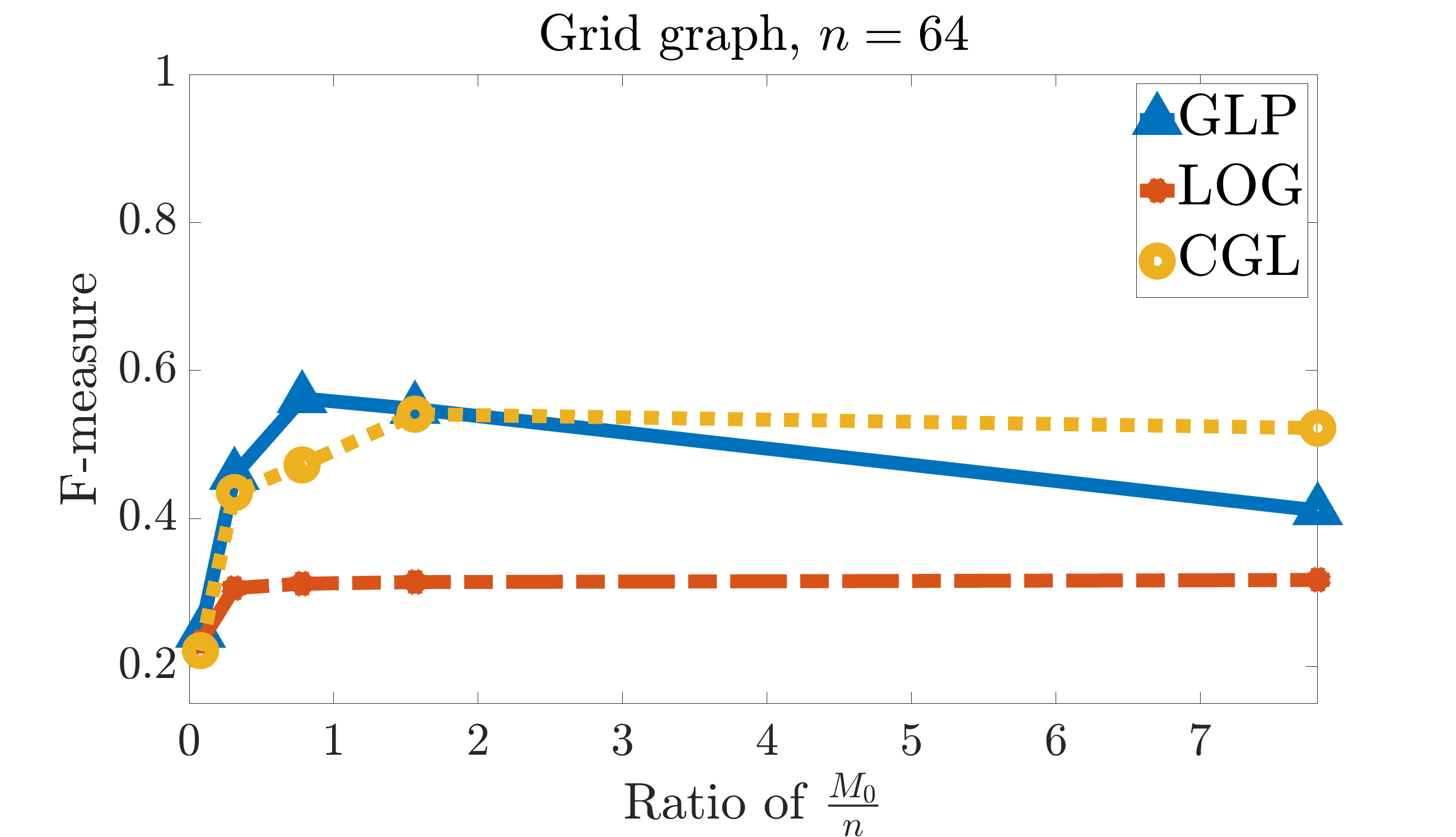}}
			{\includegraphics[height=2.6cm] {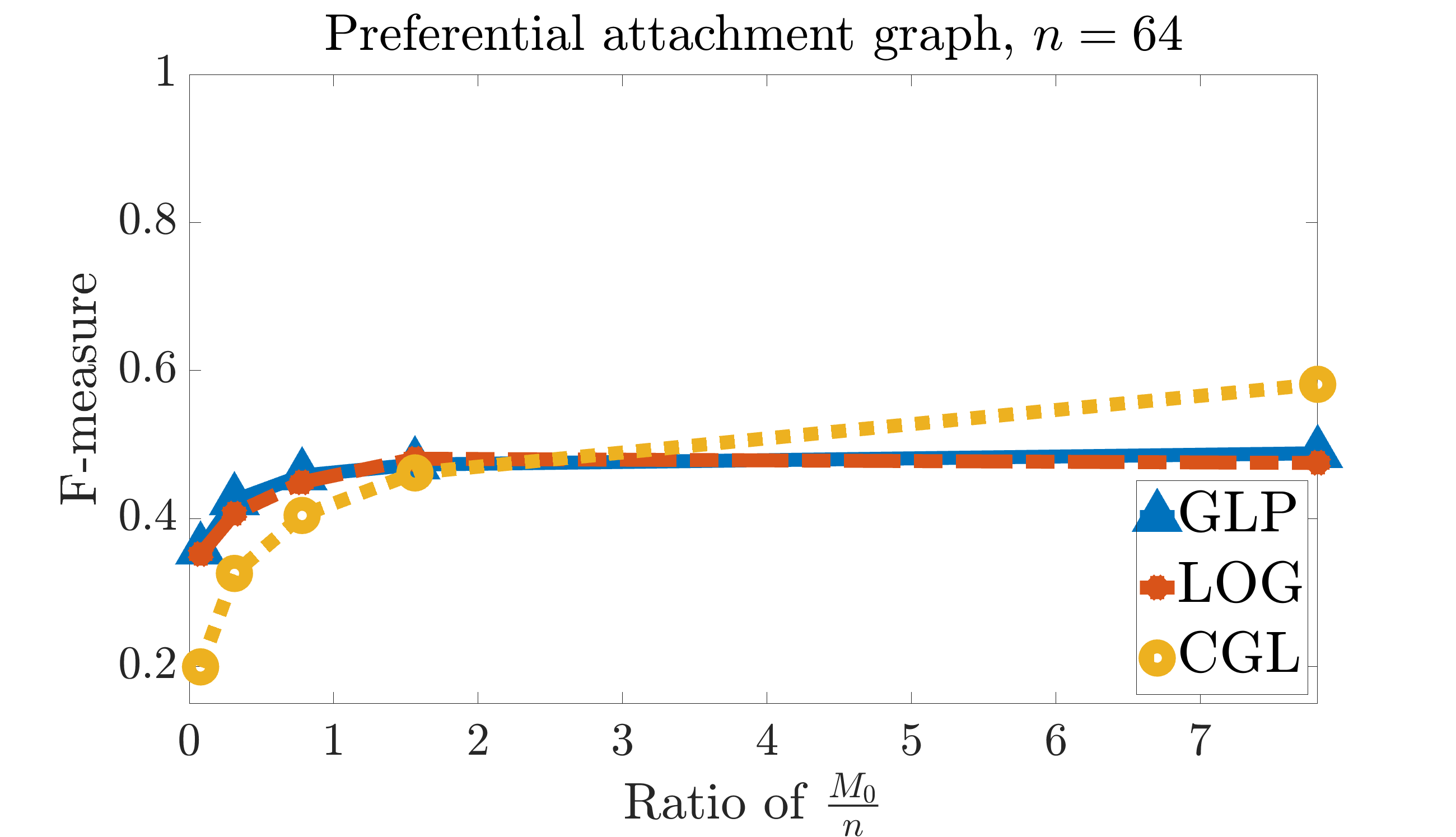}}
		\end{tabular}
	\end{center}
	\caption[example]
	%>>>> use \label inside caption to get Fig. number with \ref{}
	{ \label{fig:arbitrary:fmeas_all}
		F-measure values for various graphs for our proposed graph learning algorithm (GLP), LOG, and CGL. %GLP shines in the undersampling regime.
	}
\end{figure*}
%%%%%%%%%%%%%%%%%%%%%%%%%%%%%%%%%%%%%%%%%%%%%%%%%%%%%%%%%%%%%%%%%%%%%%%%%%%%%%%%%%%%%%%%%%%%%%%%%%%%%%%%%%%%%%%%%%%%%%%%%%%%%%%%%%%%%%%%%%%%%%%%%%%%%%
\subsection{Synthetic data: Arbitrary graphs} \label{sec:numerical_experiments:synthetic_data:arbitrary}
To showcase the performance of our new formulation for graph learning, we run synthetic experiments on a graph with $n = 64$ nodes. We generate various different graph types such as: (i) a sparse random graph with Gaussain weights, (ii) an Erdos-Renyi random graph with edge probability 0.7, (iii) a scale-free graph with preferential attachment (6 edges at each step), (iv) a random regular graph where each node is connected to $0.7n$ other nodes, (v) a uniform grid graph, (vi) a spiral graph , (vii) a community graph , and (viii) a low stretch tree graph on a grid of points. The related details of how the graphs are simulated can be found in \cite{perraudin2014gspbox,dong2016learning}. For each kind of graph, we generate 20 different realizations, and for each realization we generate observations using a degenerate multivariate Gaussian distribution with the graph Laplacian as the precision matrix \cite{dong2016learning, egilmez2017graph,kalofolias2016learn}.

We compare the performance of our proposed method with two other state-of-the-art methods for arbitrary graph learning: (i) combinatorial graph learning from \cite{egilmez2017graph} (which we refer to as CGL), and (ii) graph learning method from \cite{kalofolias2017large} (which we refer to as LOG), which also aims to learn a combinatorial graph Laplacian through a slightly different optimization problem than \cite{egilmez2017graph}. We choose $\alpha$ for our algorithm in the range $0.75^{\{0:14\}} \times \sqrt{\frac{\log(n)}{M_{0}}}$, as dictated by the error bounds for learning graphs in App.~\ref{app:th:kronecker} and by the existing literature \cite{dong2016learning,egilmez2017graph,kalofolias2016learn}. Furthermore, we choose $\rho = 0.75/\log(M_{0})$ as the value that works best for most cases. For each algorithm, in the prescribed range of the optimization parameters, we choose the parameters that produce the best results.

\begin{figure} [t]
	\begin{center}
		\begin{tabular}{cccc}
			{\includegraphics[height=2.6cm] {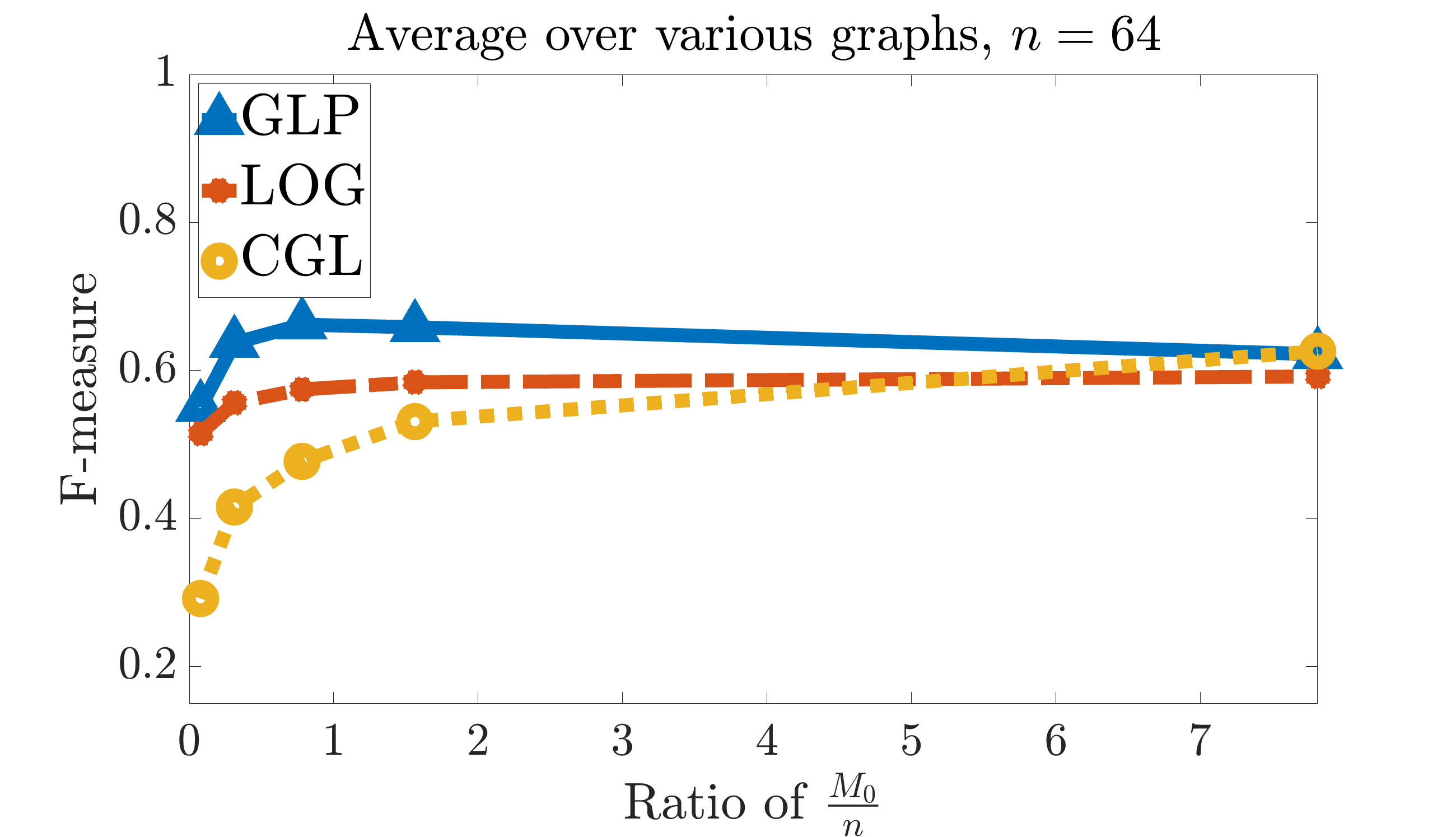}}
			{\includegraphics[height=2.6cm] {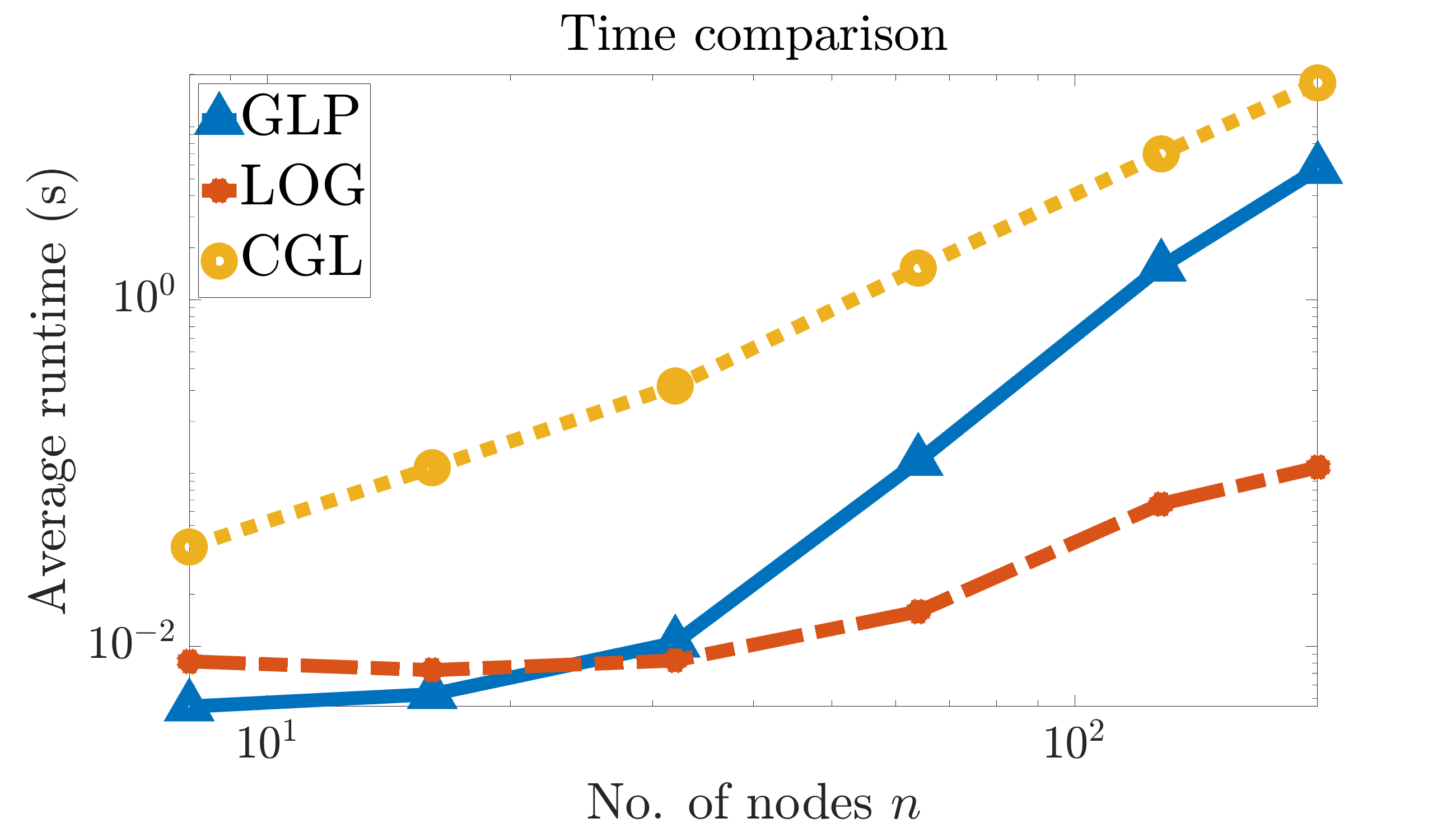}}
		\end{tabular}
	\end{center}
	\caption[example]
	%>>>> use \label inside caption to get Fig. number with \ref{}
	{ \label{fig:arbitrary:fmeas_avg}
		Average F-measure values over all graphs (left) from Fig.~\ref{fig:arbitrary:fmeas_all} for our proposed graph learning algorithm (GLP), LOG, and CGL. Average run times over 30 trials for each algorithm (right), with increasing number of nodes.}
\end{figure}

The results of our experiments are shown as F-measure values in Fig.~\ref{fig:arbitrary:fmeas_all}.  F-measure is the harmonic mean of precision and recall, and signifies the overall accuracy of the algorithm \cite{egilmez2017graph}. Precision here denotes the fraction of true graph edges recovered among all the recovered edges, and recall signifies the fraction of edges recovered from the true graph edges.
One can see that our algorithm (except for community graphs) performs just as well or better than the existing state-of-the-art algorithms. Moreover, the average performance over all graphs in Fig.~\ref{fig:arbitrary:fmeas_avg} shows that on average we outperform the existing algorithms. A runtime comparison of all algorithms in Fig.~\ref{fig:arbitrary:fmeas_avg} also reveals competitive run time for our proposed scheme. The LOG algorithm, with the smallest runtime, has a huge computational overhead incurred by the construction of a matrix of pairwise distances of all rows of data matrix $\bX$. In contrast, other algorithms work with the graph signal observations directly and do not require preprocessing steps.

\begin{remark}
For some graphs in Fig.~\ref{fig:arbitrary:fmeas_all}, the performance for GLP seems to worsen as number of observations grow. This is likely due to the limited range that we have considered for searching the optimization parameter. For a bigger range, this downward trend will disappear. The range that we have prescribed is the one mostly used in the literature and on average works well in most settings.
\end{remark}
%%%%%%%%%%%%%%%%%%%%%%%%%%%%%%%%%%%%%%%%%%%%%%%%%%%%%%%%%%%%%%%%%%%%%%%%%%%%%%%%%%%%%%%%%%%%%%%%%%%%%%%%%%%%%%%%%%%%%%%%%%%%%%%%%%%%%%%%%%%%%%%%%%%%%%
%%%%%%%%%%%%%%%%%%%%%%%%%%%%%%%%%%%%%%%%%%%%%%%%%%%%%%%%%%%%%%%%%%%%%%%%%%%%%%%%%%%%%%%%%%%%%%%%%%%%%%%%%%%%%%%%%%%%%%%%%%%%%%%%%%%%%%%%%%%%%%%%%%%%%%
\subsection{Synthetic data: Product graphs} \label{sec:numerical_experiments:synthetic_data}
\begin{figure} [h]
	\begin{center}
		\begin{tabular}{c}
			{\includegraphics[height=4.5cm] {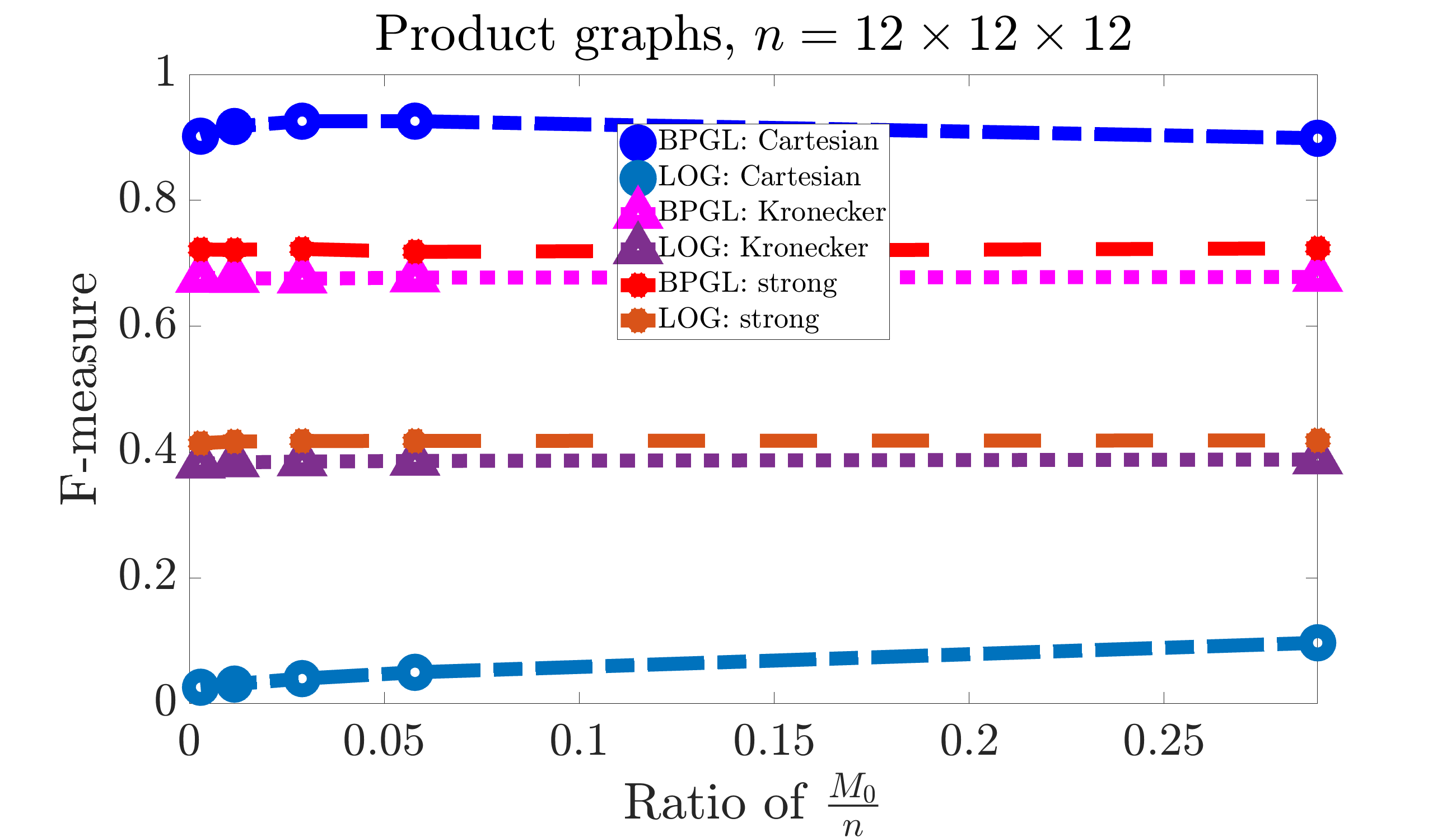}} \\
			{\includegraphics[height=4.5cm] {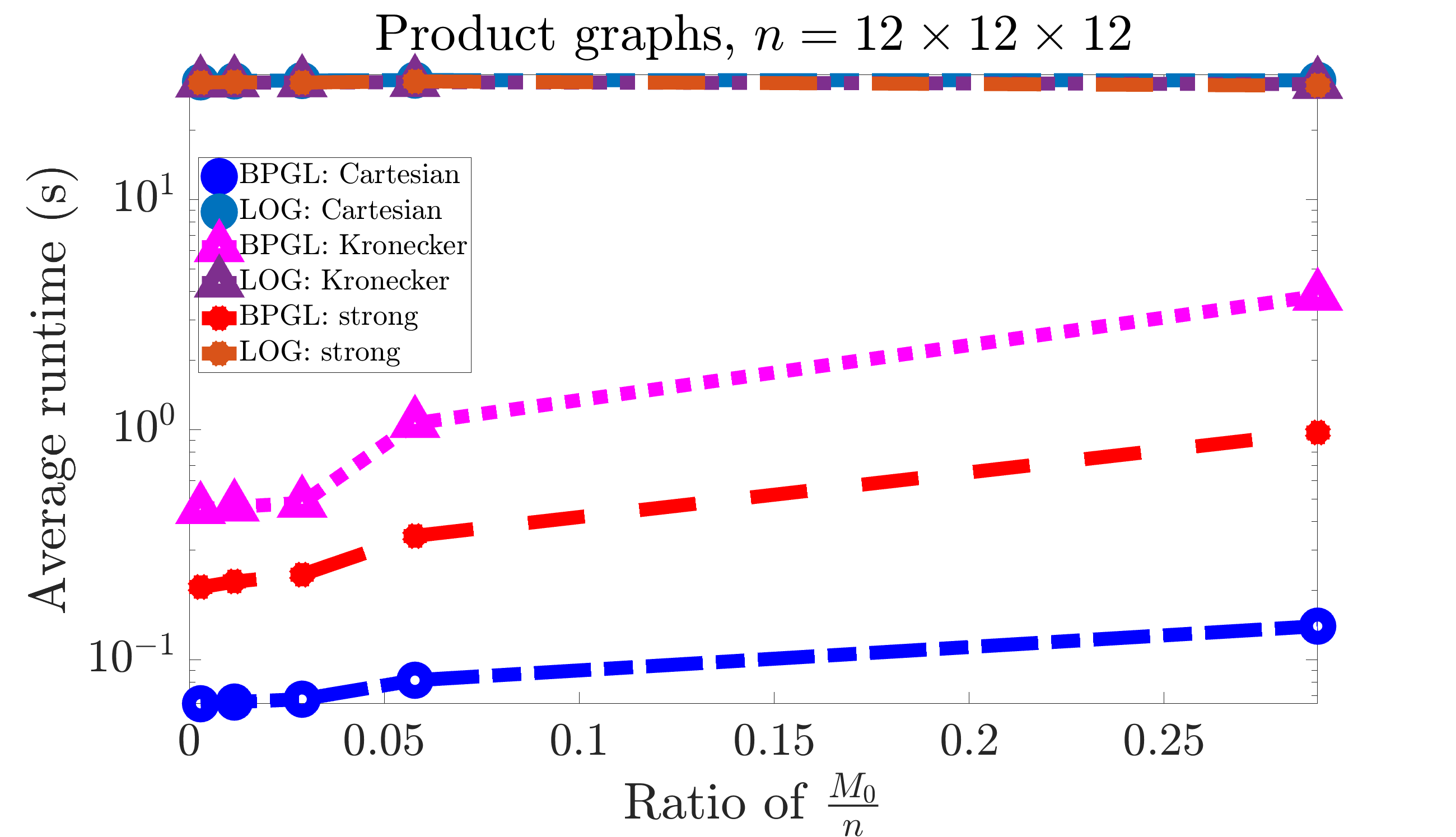}}
		\end{tabular}
	\end{center}
	\caption[example]
	%>>>> use \label inside caption to get Fig. number with \ref{}
	{ \label{fig:fmeas_run_time}
		Precision, recall and F-measure values for various values of the $\beta$ parameter. The plots shown are for Cartesian (top), Kronecker (middle), and strong (bottom) graphs when using only 5 observations for learning.}
\end{figure}

We now present the results of our numerical experiments involving synthetic data for product graphs. We run experiments for random Erdos-Renyi factor graphs with $n = n_{1} n_{2} n_{3} = 12 \times 12 \times 12$ nodes, and having either Cartesian, Kronecker or strong structure. We then use our proposed algorithms to learn the generated graphs with varying number of observations and compare the performance with LOG as its performance was the second best in Fig.~\ref{fig:arbitrary:fmeas_all}. The results for all three types of product graphs are shown in Fig.~\ref{fig:fmeas_run_time} (top). For a fixed number of observations, Cartesian product graphs can be learned with the highest F-measure score followed by strong and then Kronecker graphs. The figure also shows that for each graph, imposing product structure on the learned graph drastically improves the performance of the learning algorithm.
%In Fig.~\ref{fig:prec_rec_fmeas_obs} we also plot the F-measure scores for increasing number of observations and at various $\beta$ values. One can see that F-measure scores improve for all product graph types when more observations are used for learning.
Fig.~\ref{fig:fmeas_run_time} (bottom) also shows the runtimes comparison of our approach BPGL with the algorithm in \cite{kalofolias2017large}. Even for a graph of this size, with total number of nodes $n = 1728$, we can see a considerable reduction in runtimes. Thus, our learning algorithm that explicitly incorporates the product structure of the graph enjoys superior performance, reduced computational complexity and faster runtimes.
%%%%%%%%%%%%%%%%%%%%%%%%%%%%%%%%%%%%%%%%%%%%%%%%%%%%%%%%%%%%%%%%%%%%%%%%%%%%%%%%%%%%%%%%%%%%%%%%%%%%%%%%%%%%%%%%%%%%%%%%%%%%%%%%%%%%%%%%%%%%%%%%%%%%%%
\subsection{United States wind speed data} \label{sec:numerical_experiments:us_wind}
The first real data we use for experimentation is NCEP wind speed data.
The NCEP wind speed data \cite{kalnay1996ncep} represents wind conditions in the lower troposphere and contains daily averages of U (east-west) and V (north-south) wind components over the years $1948-2012$. Similar to the experiments in \cite{tsiligkaridis2013covariance} with preprocessed data, we use a grid of $n_{1} n_{2} = 10 \times 10 = 100$ stations to extract the data available for the United States region. From this extracted data, we choose the years $2003-2007$ for training and keep the years $2008-2012$ for testing purposes. Using a non-overlapping window of length $n_{3} = 8$, which amounts to dividing the data into chunks of 8 days, we obtain $M_{0} = 228$ samples, each of length $n = n_{1} n_{2} n_{3}$. Therefore, for graph learning we have 228 samples, where each sample contains spatiotemporal data for 100 stations over 8 days. Next, the testing procedure consists of introducing missing values in each sample of the test data by omitting the data for the 8th day, and then using a linear minimum-mean-square-error (LMMSE) estimator \cite[Chapter 4]{luenberger1997optimization} to predict the missing values. As for learning, 228 samples are obtained for testing through the same procedure.
Our proposed method estimates the (structured) adjacency matrix of the graph (which is related to the precision matrix of the data), and we use $\bW + \bI$ in place of the data covariance for the LMMSE estimator.

We make a comparison of the following approaches: (1) sample covariance matrix (SCM) learning, (2) the permuted rank-penalized least squares (PRLS) approach \cite{tsiligkaridis2013covariance} with $r = 6$ Kronecker components, (3) PRLS with $r = 2$ Kronecker components, (4) time-varying graph learning (TVGL) approach from \cite{yamada2019time} (which was shown to outperform the approach in \cite{kalofolias2017learning}), (5) spatiotemporal strong graph with BPGL with a spatial component of size $n_{1} n_{2}$ and a temporal component of size $n_{3}$, and (6) spatiotemporal Cartesian graph with BPGL of the same dimensions. The parameters for PRLS were chosen for optimal performance as given in \cite{tsiligkaridis2013covariance}. The optimization parameters for TVGL and BPGL were manually tuned for best performance. It should be noted here that SCM and PRLS aim to learn a covariance matrix and a structured covariance matrix from the data, respectively. %In contrast, our graph learning algorithm is equivalent to learning a structured precision matrix from the data. 

The SCM aims to estimate $\frac{n(n+1)}{2}$ parameters, while the number of parameters that PRLS needs to estimate is $r ( \frac{n_{1} n_{2}(n_{1} n_{2}+1)}{2} + \frac{n_{3}(n_{3}+1)}{2} )$. On the other hand, TVGL estimates $n_{3} ( \frac{n_{1} n_{2}(n_{1} n_{2}+1)}{2})$ parameters, while BPGL needs to learn only $\frac{n_{1} n_{2}(n_{1} n_{2}+1)}{2} + \frac{n_{3}(n_{3}+1)}{2}$parameters for both strong and Cartesian graphs. %, and only $( {n_{1}(n_{2} + 1) + n_{2}(n_{2} + 1)} + {n_{3} (n_{3} + 1)} )/2$ parameters for the case of three component graphs. 
The mean prediction root-mean-squared errors (RMSE) for all methods are shown in Table~\ref{table_wind}. One can see that our proposed method outperforms PRLS and TVGL while estimating far fewer parameters than both. The table also shows that learning a strong graph for this data results in a higher RMSE reduction over the baseline (SCM), and is thus better suited for this data than the Cartesian product graph.
\begin{table}
	\caption{Comparison of prediction RMSE for US wind speed data}
	\centering
	\begin{tabular}{|c|c|c|c|c|c|c|}
		\hline
		\textbf{Method} 			& {\textbf{\begin{tabular}[c]{@{}c@{}}RMSE reduction \\over SCM (dB)\end{tabular}}} & \textbf{parameters} \\
		\hline
		{SCM} 													& -- 					& 320400 \\ %0.3444
		\hline
		{TVGL \cite{yamada2019time}}							& 1.0461			  	& 40656 \\ %0.2299
		\hline
		{PRLS \cite{tsiligkaridis2013covariance}} ($r=6$) 		& 1.7780	   			& 30492 \\ %0.2287
		\hline
		{PRLS \cite{tsiligkaridis2013covariance}} ($r=2$) 		& -1.5473			 	& 10164 \\ %0.4918
		\hline
		\textbf{BPGL strong}									& \textbf{1.8640}	  	& \textbf{5082} \\ %0.2299
		\hline
		{BPGL Cartesian}										& {1.3105}				& 5082 \\ %0.2291
		\hline
	\end{tabular}
	{\label{table_wind}\\
		Comparison of our graph learning method with SCM, PRLS and TVGL. Our proposed method outperforms the existing methods for time-varying graph learning and for learning structured covariance matrix. Moreover, our proposed procedure outperforms while using considerably fewer parameters.}
\end{table}

\subsection{ABIDE fMRI data: Exploratory data analysis} \label{sec:numerical_experiments:fmri}
The second real data that we use as an application for our proposed algorithm is a part of the ABIDE fMRI dataset \cite{craddock2013neuro,narayan_2015}. Our aim is to learn the graphs over the fMRI data of control and autistic subjects and to use the learned graphs to higlight the differences in the control and autistic brains. The data we obtain is already preprocessed to remove various fMRI artifacts and for controlization of the obtained scans \cite{narayan2016mixed}. The final preprocessed data consists of measurements from $n_{1} = 111$ brain regions scanned over $116$ time instances for each subject. The data contains scans for control and autistic subjects. To avoid class imbalance we randomly choose $47$ subjects for each class. Out of the $47$ subjects for each class, we then randomly choose $30$ subjects for training and keep the remaining $17$ for testing purposes. We use a non-overlapping window length of $n_{2} = 29$ which results into $M_{0} = 120$ samples of length $n = n_{1} \times n_{2}$.

\begin{figure}
	\begin{center}
		\begin{tabular}{c c}
			{\includegraphics[height=3.8cm] {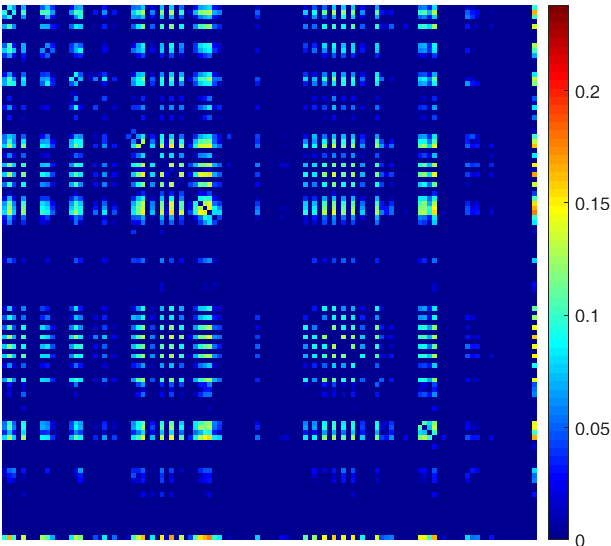}} 
			{\includegraphics[height=3.8cm] {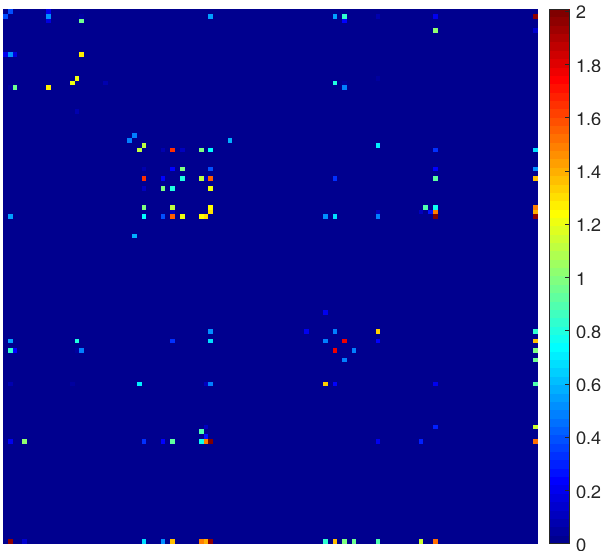}}
		\end{tabular}
	\end{center}
	\caption[example]
	%>>>> use \label inside caption to get Fig. number with \ref{}
	{ \label{fig:fmri:spatial}
		This figure shows the adjacency matrix of the spatial components learned for control (left) and autism (right) subjects with strong graph learning algorithms, respectively. The images reveal, in line with the existing literature, that control brain is much more connected than the autistic brain.}
\end{figure}

\begin{figure}
	\begin{center}
		\begin{tabular}{c c}
			{\includegraphics[height=3.4cm] {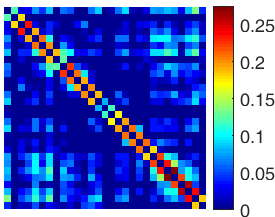}}
			{\includegraphics[height=3.4cm] {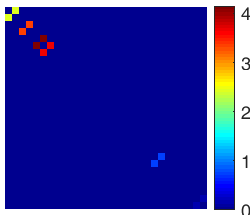}}
		\end{tabular}
	\end{center}
	\caption[example]
	%>>>> use \label inside caption to get Fig. number with \ref{}
	{ \label{fig:fmri:temporal}
		This figure shows the adjacency matrix of the temporal components learned for control (left) and autism (right) subjects with strong graph learning algorithms, respectively. The images reveal that control brains exhibit more temporal connections as compared to autistic brains. This is a new finding possible only by considering the spatiotemporal dynamics of the brain rather than just spatial connectivity analysis.}
\end{figure}

As before, we compare the performance of our proposed approach with SCM and PRLS. Table~\ref{table_fmri} shows the results of our experiments. One can see that our approach performs very similar to PRLS for both Cartesian and strong product graphs, all the while using much fewer parameters (five times fewer). We also see that strong product graphs are more suited to model brain activity. The work in \cite{narayan2016mixed} suggests that autistic brains exhibit hypoconnectivity in different regions of the brain as compared to control subjects. The results from our graph learning procedure go a step further and bring more insight into the spatiotemporal dynamics of the brain. Firstly, as already suggested in \cite{narayan2016mixed}, we see clear evidence of spatial hypoconnectivity (see Fig.~\ref{fig:fmri:spatial}). More importantly, our learned graphs in Fig.~\ref{fig:fmri:temporal} reveal that, in addition to spatial hypoconnectivity, autistic brains also suffer from temporal hypoconnectivity.

\begin{table}
	\caption{Comparison of prediction RMSE for ABIDE fMRI data}
	\centering
	\begin{tabular}{|c|c|c|c|c|c|c|}
		\hline
		\textbf{Method} 				& {\textbf{\begin{tabular}[c]{@{}c@{}}RMSE reduction \\over SCM (dB)\end{tabular}}} & \textbf{parameters} \\
		\hline
		{SCM} 												& -- 					& 5182590 \\
		\hline \hline
		\textbf{PRLS Normal}										& \textbf{2.1793} 					& 33255 \\
		\hline
		{Cartesian GL Control} 								& 2.0980				& 6651 \\
		\hline
		{Strong GL Control}							& {2.1753}		& 6651 \\
		\hline \hline
		\textbf{PRLS Autism}										& \textbf{2.375} 					& 33255 \\
		\hline
		{Cartesian GL Autism} 						& {2.3400}		& 6651 \\
		\hline
		{Strong GL Autism}									& {2.3563}				& 6651 \\
		\hline
	\end{tabular}
	{\label{table_fmri}\\
		Comparison of our graph learning method with SCM and PRLS.}
\end{table}
%%%%%%%%%%%%%%%%%%%%%%%%%%%%%%%%%%%%%%%%%%%%%%%%%%%%%%%%%%%%%%%%%%%%%%%%%%%%%%%%%%%%%%%%%%%%%%%%%%%%%%%%%%%%%%%%%%%%%%%%%%%%%%%%%%%%%%%%%%%%%%%%%%%%%%
%%%%%%%%%%%%%%%%%%%%%%%%%%%%%%%%%%%%%%%%%%%%%%%%%%%%%%%%%%%%%%%%%%%%%%%%%%%%%%%%%%%%%%%%%%%%%%%%%%%%%%%%%%%%%%%%%%%%%%%%%%%%%%%%%%%%%%%%%%%%%%%%%%%%%%
%%%%%%%%%%%%%%%%%%%%%%%%%%%%%%%%%%%%%%%%%%%%%%%%%%%%%%%%%%%%%%%%%%%%%%%%%%%%%%%%%%%%%%%%%%%%%%%%%%%%%%%%%%%%%%%%%%%%%%%%%%%%%%%%%%%%%%%%%%%%%%%%%%%%%%
%%%%%%%%%%%%%%%%%%%%%%%%%%%%%%%%%%%%%%%%%%%%%%%%%%%%%%%%%%%%%%%%%%%%%%%%%%%%%%%%%%%%%%%%%%%%%%%%%%%%%%%%%%%%%%%%%%%%%%%%%%%%%%%%%%%%%%%%%%%%%%%%%%%%%%
%%%%%%%%%%%%%%%%%%%%%%%%%%%%%%%%%%%%%%%%%%%%%%%%%%%%%%%%%%%%%%%%%%%%%%%%%%%%%%%%%%%%%%%%%%%%%%%%%%%%%%%%%%%%%%%%%%%%%%%%%%%%%%%%%%%%%%%%%%%%%%%%%%%%%%%
%\begin{figure} [ht]
%	\begin{center}
%		\begin{tabular}{c}
%			{\includegraphics[height=7cm] {spat_temporal_control_strn.png}} \\
%			{\includegraphics[height=7cm] {spat_temporal_autism_cart.png}}
%		\end{tabular}
%	\end{center}
%	\caption[example]
%	%>>>> use \label inside caption to get Fig. number with \ref{}
%	{ \label{fig:fmri:spatiotemporal}
%		This figure shows the spatiotemporal adjacency matrices for control and autism subjects with strong and cartesian learning algorithms, respectively. Normal brains have more connections and higher strength as compared to autistic brains.}
%\end{figure}
%%%%%%%%%%%%%%%%%%%%%%%%%%%%%%%%%%%%%%%%%%%%%%%%%%%%%%%%%%%%%%%%%%%%%%%%%%%%%%%%%%%%%%%%%%%%%%%%%%%%%%%%%%%%%%%%%%%%%%%%%%%%%%%%%%%%%%%%%%%%%%%%%%%%%%%
\subsection{Estrogen receptor data} \label{sec:numerical_experiments:er}
The final dataset that we experiment on is the estrogen receptor data \cite{dobra2009variable,li2010inexact}, which consists of 157 samples of 693 probe sets related to estrogen receptor pathway. %(after selecting the appropriate relevant subset of probes). 
We aim to learn a Kronecker structured graph on this data using $120$ randomly selected samples for training and the remaining $37$ for testing. We choose Kronecker structured graph for this data because transposable models, i.e., models that learn a Kronecker structured data covariance, have been shown to work well for genomic data in the existing literature \cite{allen2010transposable}. And as pointed out in Sec.~\ref{sec:product_graph_prob:kron}, Kronecker structured adjacency matrix corresponds to a Kronecker structured data covariance.
For testing purposes, we follow a procedure similar to the previous subsections.% where we remove part of data from test samples, $33$ probe set measurements in this case, and use linear MMSE estimator to complete the data. 

We compare our graph learning approach with SCM, PRLS and sparse covariance estimation (SEC) from \cite{cui2016sparse}. Optimization parameters are manually tuned for best results for each method. We learn a graph through our method (and covariance through PRLS), with a Kronecker structure composed of two factor matrices of dimensions $n_{1} = 21$ and $n_{2} = 33$. We then use LMMSE estimator to predict $33$ probe set measurements removed from the test data. PRLS, SEC and BPGL result in an improvement of $\mathbf{0.91347}$ dB, $\mathbf{0.93598}$ dB, and $\mathbf{1.0242}$ dB over SCM, respectively. This demonstrates that our method outperforms the state-of-the-art unstructured and structured sparse covariance estimation techniques, and provides a better model for real datasets.

%% file: conclusion.tex
\section{Conclusion}\label{sec:conc}
In this paper, we introduced a new linear formulation of the graph learning problem from graph signals. We demonstrated the performance of this new formulation with numerical experiments and derived bounds on its estimation performance.
Based on the proposed formulation, we also posed the problem to learn product graphs from data. We devised a block coordinate descent based algorithm for learning product graphs, and derived the associated error bounds for various product structures. Finally, we validated the performance characteristics and superior learning capabilities of our proposed method through numerical simulations on synthetic and real datasets.

%% file: appendix.tex
\begin{appendices} \label{app}

\section{Proof of Theorem~\ref{th:kronecker}}\label{app:th:kronecker}
Note that the current form of the objective function \eqref{learn_kronecker} can be expressed as
\begin{align} \label{eq:smoothness:kronecker}
&\frac{\alpha}{M_{0}}~\underset{m = 1}{\overset{M_{0}}{\sum}} \bar{\bx}_{m}^{T} (\underset{[K_{0}]}{\otimes} \bW_{i}) \bone - \bx_{m}^{T} (\underset{[K_{0}]}{\otimes} \bW_{i}) \bx_{m} \nonumber\\
&= \frac{\alpha}{M_{0}}~\underset{m = 1}{\overset{M_{0}}{\sum}} \langle \bar{\cX}_{m},\one \underset{[K_{0}]}{\times} \bW_{i} \rangle - \langle \cX_{m},\cX_{m} \underset{[K_{0}]}{\times} \bW_{i} \rangle
\end{align}
using the properties of the Kronecker product. Let us define the following: the tensor $\cY_{m} = \cX_{m} \underset{[K_{0}] \setminus k}{\times} \bW_{i}^{1/2} \underset{k}{\times} \bI_{k}$, the matrix $\bS_{k} = \frac{1}{M_{0}}\sum_{m=1}^{M_{0}} \cY_{m(k)} \cY_{m(k)}^{T}$, $\bar{\bd}_{k}$ as a vector of weighted degrees of the product adjacency matrix $\underset{[K_{0}] \setminus k}{\otimes} \bW_{i}$, $\bd_{k}$ as the vector of weighted degrees of $\bW_{k}$, $\bar{\by}_{mj}$ as the $j$-th column of $\bar{\cX}_{m(k)}$, $\by_{mj}$ as the $j$-th column of $\cX_{m(k)}$, and finally the matrix $\bar{\bS}_{k} = \frac{1}{M_{0}}\sum_{m=1}^{M_{0}}\sum_{j=1}^{n/n_{k}} (\bar{\bd}_{k})_{j} \by_{mj} \by_{mj}^{T}$. Then we can further express the terms in \eqref{eq:smoothness:kronecker} as:
\begin{align} \label{eq:smoothness:kronecker:1}
& \frac{\alpha}{M_{0}}~\underset{m = 1}{\overset{M_{0}}{\sum}} \langle \cX_{m},\cX_{m} \underset{[K_{0}]}{\times} \bW_{i} \rangle \nonumber\\
&= \frac{\alpha}{M_{0}}~\underset{m = 1}{\overset{M_{0}}{\sum}} \tr \big( \bW_{k} \cX_{m(k)} (\underset{[K_{0}] \setminus k}{\otimes} \bW_{i}) \cX_{m(k)}^{T} \big) \nonumber \\
&= \frac{\alpha}{M_{0}}~\underset{m = 1}{\overset{M_{0}}{\sum}} \tr \big( \bW_{k} \cY_{m(k)} \cY_{m(k)}^{T} \big) = \alpha~\tr\big( \bW_{k} \bS_{k} \big),
%&= \frac{\alpha}{M_{0}}~\underset{m = 1}{\overset{M_{0}}{\sum}} \by_{m}^{T} \bL_{k} \by_{m} \nonumber \\
%&= \alpha~ \tr\big( \bL_{k} \bS_{k} \big) \nonumber \\
%&= \frac{\alpha}{M_{0}}~\underset{m = 1}{\overset{M_{0}}{\sum}} \bar{\by}_{m}^{T} \bW_{k} \bone - \by_{m}^{T} \bW_{k} \by_{m}
\end{align}
and,
\begin{align} \label{eq:smoothness:kronecker:2}
& \frac{\alpha}{M_{0}}~\underset{m = 1}{\overset{M_{0}}{\sum}} \langle \bar{\cX}_{m},\one \underset{[K_{0}]}{\times} \bW_{i} \rangle \nonumber\\
&= \frac{\alpha}{M_{0}}~\underset{m = 1}{\overset{M_{0}}{\sum}} \tr \big( \bW_{k} \bar{\cX}_{m(k)} (\underset{[K_{0}] \setminus k}{\otimes} \bW_{i}) \one_{(k)}^{T} \big) \nonumber \\
&= \frac{\alpha}{M_{0}}~\underset{m = 1}{\overset{M_{0}}{\sum}} \tr \big( \bW_{k} \bar{\cX}_{m(k)} \bar{\bd}_{k} \bone^{T} \big) = \frac{\alpha}{M_{0}}~\underset{m = 1}{\overset{M_{0}}{\sum}} \tr \big( \bd_{k}^{T} \bar{\cX}_{m(k)} \bar{\bd}_{k} \big) \nonumber \\
&= \frac{\alpha}{M_{0}}~\underset{m = 1}{\overset{M_{0}}{\sum}} \tr \big( \bd_{k}^{T} [\bar{\by}_{m1}, \bar{\by}_{m2}, \cdots, \bar{\by}_{mn/n_{k}}] \bar{\bd}_{k} \big) \nonumber \\
&= \frac{\alpha}{M_{0}}~\underset{m = 1}{\overset{M_{0}}{\sum}}~\underset{j = 1}{\overset{n/n_{k}}{\sum}} \tr \big( (\bar{\bd}_{k})_{j} \by_{mj}^{T} \bD_{k} \by_{mj} \big) = \alpha~\tr(\bD_{k} \bar{\bS}_{k})
%&= \frac{\alpha}{M_{0}}~\underset{m = 1}{\overset{M_{0}}{\sum}}~\underset{j = 1}{\overset{n_{k}}{\sum}} (\bd_{k})_{j} \by_{m,i}^{T} \bar{\bD}_{k} \by_{m,i} \nonumber \\
%&= \frac{\alpha}{M_{0}}~\underset{m = 1}{\overset{M_{0}}{\sum}}~\underset{j = 1}{\overset{n_{k}}{\sum}} (\bd_{k})_{j} \by_{j}^{T} \bar{\bD}_{k} \by_{j},
\end{align}
Moreover, for the terms in \eqref{eq:smoothness:kronecker:1} and \eqref{eq:smoothness:kronecker:2}, the difference from their expected values takes the form of:
\begin{align} \label{eq:smoothness:kronecker:3}
&\alpha~\Big|\tr\big( \bW_{k} \bS_{k} \big) - \E_{\bx} \big[ \tr\big( \bW_{k} \bS_{k} \big) \big]\Big| \nonumber\\
&= \alpha~\Big| \tr\big( \bW_{k} \bS_{k} \big) - \tr\big( \bW_{k} \E_{\bx}[\bS_{k}]\big) \Big| \nonumber\\
&= \alpha~\Big| \tr\big( \bW_{k} (\bS_{k} - \E_{\bx}[\bS_{k}]) \big) \Big| \nonumber\\
&= \alpha~\Big| \underset{i,j}{\overset{ }{\sum}} \big( \bW_{k} (\bS_{k} - \E_{\bx}[\bS_{k}]) \big)_{i,j} \Big| \nonumber\\
&\leq \alpha~\underset{i,j}{\max}\Big|(\bS_{k} - \E_{\bx}[\bS_{k}])_{i,j}\Big|~\underset{i,j}{\overset{ }{\sum}} \big( \bW_{k} \big)_{i,j} \nonumber\\
&= \alpha n_{k}~\underset{i,j}{\max}\Big|(\bS_{k} - \E_{\bx}[\bS_{k}])_{i,j}\Big| \nonumber\\
&\leq C_{1} \frac{n_{k}^{2}\log(n_{k})}{nM_{0}} \Big({1+\big\|\underset{[K_{0}] \setminus k}{\otimes} \whbW_{i} - \underset{[K_{0}] \setminus k}{\otimes} \bW_{i}^{*} \big\|_{F}} \Big),
\end{align}
and
\begin{align} \label{eq:smoothness:kronecker:4}
&\alpha~\Big|\tr\big( \bD_{k} \bar{\bS}_{k} \big) - \E_{\bx} \big[ \tr\big( \bD_{k} \bar{\bS}_{k} \big) \big]\Big| \nonumber\\
&\leq \alpha~\underset{i,j}{\max}\Big|(\bar{\bS}_{k} - \E_{\bx}[\bar{\bS}_{k}])_{i,j}\Big|~\underset{i,j}{\overset{ }{\sum}} \big( \bD_{k} \big)_{i,j} \nonumber\\
&= \alpha n_{k}~\underset{i,j}{\max}\Big|(\bar{\bS}_{k} - \E_{\bx}[\bar{\bS}_{k}])_{i,j}\Big| \nonumber\\
&\leq C_{2} \frac{n_{k}^{2} \log(n_{k}) }{nM_{0}} \Big({1+\big\|\underset{[K_{0}] \setminus k}{\otimes} \whbW_{i} - \underset{[K_{0}] \setminus k}{\otimes} \bW_{i}^{*} \big\|_{F}} \Big),
\end{align}
with probability for both inequalities exceeding $1-4n_{k}^2 \Big[\exp\big(\frac{-nM_{0}}{2n_{k}}\big) + \exp\big(-(0.25+\sqrt{\log n_{k}})^2\big)\Big]$; details of the last inequalities of \eqref{eq:smoothness:kronecker:3} and \eqref{eq:smoothness:kronecker:4} in \cite[Lemma B.1]{sun2015non}. Here $\whbW_{i}$ is the current estimate of the $i$-th adjacency matrix while $\bW_{i}^{*}$ is the original $i$-th generating factor. The last inequalities in both expressions follow from \cite{sun2015non} by the fact that for estimating the $k$-th factor the estimates $\whbW_{i}$ for other factors are used, and by choosing $\alpha \leq \sqrt{\frac{n_{k}\log(n_{k})}{nM_{0}}}$.
With these bounds, the error between the sample-based objective and the population-based objective, which are defined as $\tr\big( \bW_{k} \bS_{k} \big) - \tr\big( \bD_{k} \bar{\bS}_{k} \big)$ with $M_{0}$ samples and as $\E_{\bx} \big[ \tr\big( \bW_{k} \bS_{k} \big) \big] + \E_{\bx} \big[ \tr\big( \bD_{k} \bar{\bS}_{k} \big) \big]$ with infinitely many samples, respectively, can be upper bounded as:
\begin{align}
&\alpha~\Big|\tr\big( \bW_{k} \bS_{k} \big) - \tr\big( \bD_{k} \bar{\bS}_{k} \big) \nonumber\\
&- \E_{\bx} \big[ \tr\big( \bW_{k} \bS_{k} \big) \big] + \E_{\bx} \big[ \tr\big( \bD_{k} \bar{\bS}_{k} \big) \big]\Big| \nonumber\\
&\leq \alpha~\Big|\tr\big( \bW_{k} \bS_{k} \big) - \E_{\bx} \big[ \tr\big( \bW_{k} \bS_{k} \big) \big]\Big| + \nonumber\\
& + \alpha~\Big|\tr\big( \bD_{k} \bar{\bS}_{k} \big) - \E_{\bx} \big[ \tr\big( \bD_{k} \bar{\bS}_{k} \big) \big]\Big| \nonumber\\
&\leq C \frac{n_{k}^{2}\log(n_{k})}{nM_{0}} \Big({1+\big\|\underset{[K_{0}] \setminus k}{\otimes} \whbW_{i} - \underset{[K_{0}] \setminus k}{\otimes} \bW_{i}^{*} \big\|_{F}} \Big).
\end{align}

To derive the bound on the error between the factor estimate $\whbW_{k}$ and the original generating $\bW_{k}^{*}$, let us first define the following convex function of $\bDelta$:
\begin{align} \label{eq:app:th:kronecker:FDelta}
F_{k}(\bDelta) &= \alpha~\tr\big( \bar{\bS}_{k} \diag(\bDelta \bone) \big) - \alpha~\tr\big( \bS_{k} \bDelta \big),
%&= \alpha \tr\big( \bD_{k}' \bar{\bS}_{k} - \bW_{k}' \bS_{k} \big) -  \tr\big( \bD_{k} \bar{\bS}_{k} - \bW_{k} \bS_{k} \big),
%&= \alpha~\tr\big( \bS_{k} \bDelta \big) \nonumber\\
%&= \alpha~\tr\big( \bS_{k} (\bL_{k} - \bL_{k}^{*}) \big) \nonumber\\
%&= \alpha~\tr\big( \bL_{k} \bS_{k} \big) - \alpha~\tr\big( \bL_{k}^{*} \bS_{k} \big),
\end{align}
where $\bDelta = \bW_{k} - \bW_{k}^{*}$, and $\diag(\bDelta \bone) = \bD_{k} - \bD_{k}^{*}$. %, and $\tr(\cdot)$ represents the trace of the argument after $\bL$ has been projected to the set of valid Laplacians.
Now, we want to prove that $F_{k}(\bDelta) > 0$, for $\bDelta \in \R^{n_{k}\times n_{k}}$ with $\|\bDelta\|_{F} = \|\bW_{k} - \bW_{k}^{*}\|_{F} = R\sqrt{\frac{n_{k}\log(n_{k})}{nM_{0}}}$ with a constant $R > 0$. %C\sqrt{\frac{n_{k}\log(n_{k})}{M_{0}\prod_{i, i\neq k}^{K_{0}} n_{i}}}
Consider $F_{k}(\cdot)$ at $\whbDelta = \whbW_{k} - \bW_{k}^{*}$, which is the minima of $F_{k}(\bDelta)$ because $\whbW_{k}$ is the minima of our factor-wise minimization of \eqref{learn_kronecker}. Then we have:
\begin{align}
F_{k} (\whbDelta) &= \alpha \tr\big( \whbD_{k} \bar{\bS}_{k} - \whbW_{k} \bS_{k} \big) -  \alpha \tr\big( \bD_{k}^{*} \bar{\bS}_{k} - \bW_{k}^{*} \bS_{k} \big), \nonumber\\
&\leq F_{k} (\bzero) = \alpha \tr\big( \bD_{k}^{*} \bar{\bS}_{k} - \bW_{k}^{*} \bS_{k} \big) - \nonumber\\
&\quad\quad\quad\quad\quad\quad \alpha \tr\big( \bD_{k}^{*} \bar{\bS}_{k} - \bW_{k}^{*} \bS_{k} \big) = 0,
\end{align}
%where $\whbW_{k}$ is the solution that minimizes our cost function under the adjacency matrix constraints. 
If we can prove that $F_{k}(\bDelta) > 0$ for $\bDelta \in \R^{n_{k}\times n_{k}}$ with a certain norm, then since $F_{k} (\whbDelta) \leq 0$, it must satisfy $\|\whbDelta\|_{F} < R\sqrt{\frac{n_{k}\log(n_{k})}{nM_{0}}}$. 
To see that $F_{k} > 0$ for $\bDelta \in \R^{n_{k}\times n_{k}}$ with the prescribed norm, first consider the following using the property of the trace of product of matrices \cite{fang1994inequalities}:
\begin{align}
&\tr\big( \bar{\bS}_{k} \diag(\bDelta \bone) \big) \geq \lambda_{n_{k}}(\bar{\bS}_{k}) \tr\big( \diag(\bDelta \bone) \big) \nonumber\\
&\quad\quad\geq \lambda_{n_{k}}(\bar{\bS}_{k}) \|\diag(\bDelta \bone) \|_{F} = \lambda_{n_{k}}(\bar{\bS}_{k}) \|\bDelta \bone \|_{F} > 0
\end{align}
since $\|\bDelta \|_{F} > 0$, and where $\lambda_{n_{k}}(\bar{\bS}_{k})$ is the minimum eigenvalue of $\bS_{k}$.
Secondly, one can also see that: %$\tr\big( \bS_{k} \bDelta \big) \leq \lambda_{1}(\bS_{k}) \tr\big( \bDelta \big) = 0$,
\begin{align}
\tr\big( \bS_{k} \bDelta \big) &\leq \lambda_{1}(\bS_{k}) \tr\big( \bDelta \big) = 0,
\end{align}
where $\lambda_{1}(\bS_{k})$ is the largest eigenvalue of $\bS_{k}$, and $\tr\big( \bDelta \big) = 0$ because of the adjacency constraints. Using the upper and lower bounds on the trace terms in \eqref{eq:app:th:kronecker:FDelta} one can see that $F_{k}(\bDelta) > 0$ which completes the proof. \qed

\section{Proof of Theorem~\ref{th:cartesian}}\label{app:th:cartesian}
Let us focus on the Cartesian objective function in \eqref{learn_cartesian}:
\begin{align} \label{eq:smoothness:cartesian}
&\frac{\alpha}{M_{0}}~\underset{m = 1}{\overset{M_{0}}{\sum}} \bx_{m}^{T} (\underset{[K_{0}]}{\oplus} \bW_{i}) \bx_{m} \nonumber\\
&= \frac{\alpha}{M_{0}}~\underset{m = 1}{\overset{M_{0}}{\sum}} \underset{k = 1}{\overset{K_{0}}{\sum}} \bx_{m}^{T} ( (\underset{[k-1]}{\otimes} \bI_{i}) \otimes \bW_{k} \otimes (\underset{[K_{0}] \setminus [k]}{\otimes} \bI_{j}) ) \bx_{m}\nonumber\\
&= \frac{\alpha}{M_{0}}~\underset{m = 1}{\overset{M_{0}}{\sum}} \underset{k = 1}{\overset{K_{0}}{\sum}} \langle \cX_{m},\cX_{m} \underset{k}{\times} \bW_{k} \rangle \nonumber\\
&= \frac{\alpha}{M_{0}} \underset{m = 1}{\overset{M_{0}}{\sum}} \underset{k = 1}{\overset{K_{0}}{\sum}} \tr \big( \bW_{k} \cX_{m(k)} \cX_{m(k)}^{T} \big) = \alpha \underset{k = 1}{\overset{K_{0}}{\sum}} \tr\big( \bW_{k} \bT_{k} \big),
\end{align}
where the matrix $\bT_{k} = \frac{1}{M_{0}}\sum_{m=1}^{M_{0}} \cX_{m(k)} \cX_{m(k)}^{T}$. Similar steps along the lines of \eqref{eq:smoothness:kronecker:2} can be followed to arrive at $\alpha/M_{0} \sum_{m=1}^{M_{0}} \bx_{m}^{T} (\underset{[K_{0}]}{\oplus} \bD_{i}) \bx_{m} = \alpha \sum_{k=1}^{K_{0}} \tr\big( \bD_{k} \bar{\bT}_{k} \big)$. Thus, one can clearly see that the objective function can be expressed as a sum of terms each of which is dependent on only one of the factor adjacency matrices $\bW_{k}$. As in Appendix~\ref{app:th:kronecker}, using \cite[Lemma B.1]{sun2015non} we can see that with high probability:
\begin{align}
&\underset{i,j}{\max}\Big|(\bT_{k} - \E_{\bx}[\bT_{k}])_{i,j}\Big| \leq C_{1} \frac{n_{k}^{2}\log(n_{k})}{nM_{0}},
\end{align}
since each term in the objective function \eqref{eq:smoothness:cartesian} is dependent on only one factor adjacency matrix.
After this, one can follow the steps in Appendix~\ref{app:th:kronecker} to obtain the final error bounds. \qed

\section{Proof of Theorem~\ref{th:strong}}\label{app:th:strong}
Focusing again on the objective in \eqref{learn_strong}, the terms involving only the $k$-th factor $\bL_{k}$ can be expressed as: 
\begin{align} \label{eq:smoothness:strong}
&\frac{\alpha}{M_{0}}~\sum_{m = 1}^{M_0} \sum_{j = 0}^{K_{0}-1} \sum_{\bp}^{\bP(k,j)}  \cX_{m} {:} (\cX_{m} \times_{\bp} \bW_{\bp} \times_{k} \bW_{k}) \nonumber\\
&= \frac{\alpha}{M_{0}}~\sum_{m = 1}^{M_0} \sum_{j = 0}^{K_{0}-1} \sum_{\bp}^{\bP(k,j)} \tr \big( \bW_{k} \cX_{m(k)} \bM_{\bp} \cX_{m(k)}^{T} \big) \nonumber \\
&= \frac{\alpha}{M_{0}}~\sum_{m = 1}^{M_0} \tr \big( \bW_{k} \cX_{m(k)} \Big[ \sum_{j = 0}^{K_{0}-1} \sum_{\bp \in \bP(k,j)} \bM_{\bp} \Big] \cX_{m(k)}^{T} \big) \nonumber \\
&= \frac{\alpha}{M_{0}}~\sum_{m = 1}^{M_0} \tr \big( \bW_{k} \cX_{m(k)} \bQ_{k} \cX_{m(k)}^{T} \big) = \alpha~\tr\big( \bW_{k} \bZ_{k} \big),
\end{align}
where $\bp$ denotes a column from matrix $\bP(k,j)$, $\sum_{\bp \in \bP(k,j)}$ denotes the summation over the columns of $\bP(k,j)$, and the columns of $\bP(k,j)$ are different combinations of indices given by $C^{[1,\dots,K_0]\setminus[k]}_{j}$. Additionally, $\bM_{\bp}$ denotes a matrix of size $(n-n_{k}) \times (n-n_{k})$ that contains an appropriate Kronecker product of identity matrices and factor adjacency matrices in accordance with the entries of the vector $\bp$, and $\bZ_{k} = \frac{1}{M_{0}}\sum_{m = 1}^{M_0} \cX_{m(k)} \bQ_{k} \cX_{m(k)}^{T}$.
%The remaining steps are similar to the proof of Theorem~\ref{th:cartesian}, and are thus omitted in the interest of space. 
Expressing the objective as a sum of terms that all contain the $k$-th factor facilitates the process of solving factor-wise problems.

To derive the error bounds see that the objective function in \eqref{learn_strong} can be expressed as: $\frac{\alpha}{M_{0}}~\underset{m = 1}{\overset{M_{0}}{\sum}} \bar{\bx}_{m}^{T} (\underset{[K_{0}]}{\boxtimes} \bW_{i}) \bone - \bx_{m}^{T} (\underset{[K_{0}]}{\boxtimes} \bW_{i}) \bx_{m}$.
%\begin{align} \label{eq:smoothness:strong}
%\frac{\alpha}{M_{0}}~\underset{m = 1}{\overset{M_{0}}{\sum}} \bar{\bx}_{m}^{T} (\underset{[K_{0}]}{\boxtimes} \bW_{i}) \bone - \bx_{m}^{T} (\underset{[K_{0}]}{\boxtimes} \bW_{i}) \bx_{m}.
%\end{align}
The difference of the second term in this expression from its expected value can be further expressed and upper bounded as:
\begin{align}
&\frac{\alpha}{M_{0}} \Big|\underset{m = 1}{\overset{M_{0}}{\sum}} \bx_{m}^{T} (\underset{[K_{0}]}{\boxtimes} \bW_{i}) \bx_{m} - \E_{\bx} \big[ \bx_{m}^{T} (\underset{[K_{0}]}{\boxtimes} \bW_{i}) \bx_{m} \big]\Big| \nonumber \\
&=\frac{\alpha}{M_{0}} \Big|\tr \Big( (\underset{[K_{0}]}{\boxtimes} \bW_{i}) \underset{m = 1}{\overset{M_{0}}{\sum}} \bx_{m}^{T}  \bx_{m}\Big) - \nonumber \\
&\quad\quad\quad\quad\quad\quad \tr \Big( (\underset{[K_{0}]}{\boxtimes} \bW_{i}) \E_{\bx} \big[\underset{m = 1}{\overset{M_{0}}{\sum}} \bx_{m}^{T}  \bx_{m}\big] \Big) \Big| \nonumber \\
&\leq \frac{\alpha \lambda_{1}(\bW_{k})}{M_{0}} \Big|\tr \Big( ( \underset{[k-1]}{\boxtimes} \bW_{i} \boxtimes \bI_{k} \underset{[K_{0}] \setminus [k]}{\boxtimes} \bW_{j} ) \underset{m = 1}{\overset{M_{0}}{\sum}} \bx_{m}^{T}  \bx_{m}\Big) \nonumber \\
&\quad\quad - \tr \Big( ( \underset{[k-1]}{\boxtimes} \bW_{i} \boxtimes \bI_{k} \underset{[K_{0}] \setminus [k]}{\boxtimes} \bW_{j} ) \E_{\bx} \big[\underset{m = 1}{\overset{M_{0}}{\sum}} \bx_{m}^{T}  \bx_{m}\big] \Big) \Big| \nonumber \\
&\leq C_{1} \frac{n_{k}\log(n)}{M_{0}} \Big({1+2\big\|\underset{[K_{0},k]}{\boxtimes} \whbW_{i} - \underset{[K_{0}]}{\boxtimes} \bW_{i}^{*} \big\|_{F}} \Big),
\end{align}
where $\underset{[K_{0},k]}{\boxtimes} \whbW_{i} = \underset{[k-1]}{\boxtimes} \bW_{i} \boxtimes \bI_{k} \underset{[K_{0}] \setminus [k]}{\boxtimes} \bW_{j}$, and the last inequality follows with high probability from \cite[Lemma B.1]{sun2015non}.  With this bound, the remaining steps are similar to the proof of Theorem~\ref{th:cartesian}, and are thus omitted in the interest of space. \qed

\section{Proof of Theorem~\ref{th:convergence}}\label{app:th:convergence}
To prove this theorem, %one can follow along the lines of the proof for \cite[Proposition~2.7.1]{bertsekas1999nonlinear}. 
we first realize that since each mode-wise problem is convex, the update for each mode-wise problem is guaranteed to converge to its minimum. %Once block/mode-wise convergence to the minima is established, the convergence of every limit point to a stationary point is proven from \cite[Proposition~2.7.1]{bertsekas1999nonlinear}.
The rate of convergence can then be established through the work in \cite{xu2013block} which provides convergence guarantees and rates of convergence for block coordinate descent for multiconvex objectives. It can be trivially seen that each factor-wise problem \eqref{eq:lagrangian} for learning factor graphs is strongly convex. The strong convexity of the factor-wise problems, in conjunction with \cite[Theorem~2.9]{xu2013block}, %(part 2), 
implies that Alg.~\ref{alg_learn_product_graph} presented in this paper converges to its critical points at a linear rate.\qed

\end{appendices}